\documentclass[twocolumn]{aastex631}

\received{May 14, 2021}
\revised{August 12, 2021}
\accepted{August 16, 2021}
\submitjournal{AAS Journals}

\shorttitle{Metal-poor Stars in the Magellanic Clouds}
\shortauthors{Reggiani et al.}

\begin{document}

\title{The Most Metal-poor Stars in the Magellanic Clouds are $r$-process
Enhanced\footnote{This paper includes data gathered with the 6.5-meter
Magellan Telescopes located at Las Campanas Observatory, Chile.}}

\correspondingauthor{Henrique Reggiani}
\email{hreggiani@jhu.edu}

\author[0000-0001-6533-6179]{Henrique Reggiani}
\affiliation{Department of Physics \& Astronomy, Johns Hopkins
University, 3400 N Charles St., Baltimore, MD 21218, USA}

\author[0000-0001-5761-6779]{Kevin C.\ Schlaufman}
\affiliation{Department of Physics \& Astronomy, Johns Hopkins
University, 3400 N Charles St., Baltimore, MD 21218, USA}
\affiliation{Center for Computational Astrophysics, Flatiron Institute,
162 5th Ave, New York, NY 10010, USA}

\author[0000-0003-0174-0564]{Andrew R.\ Casey}
\affiliation{School of Physics \& Astronomy, Monash University, Wellington
Road, Clayton 3800, Victoria, Australia}
\affiliation{ARC Centre of Excellence for All Sky Astrophysics in 3
Dimensions(ASTRO 3D), Canberra, ACT 2611, Australia}

\author[0000-0002-4733-4994]{Joshua D.\ Simon}
\affiliation{The Observatories of the Carnegie Institution for Science,
813 Santa Barbara St., Pasadena, CA 91101, USA}

\author[0000-0002-4863-8842]{Alexander P.\ Ji}
\altaffiliation{Hubble Fellow}
\affiliation{The Observatories of the Carnegie Institution for Science,
813 Santa Barbara St., Pasadena, CA 91101, USA}

\begin{abstract}

\noindent
The chemical abundances of a galaxy's metal-poor stellar population can
be used to investigate the earliest stages of its formation and chemical
evolution.  The Magellanic Clouds are the most massive of the Milky
Way's satellite galaxies and are thought to have evolved in isolation
until their recent accretion by the Milky Way.	Unlike the Milky Way's
less massive satellites, little is know about the Magellanic Clouds'
metal-poor stars.  We have used the mid-infrared metal-poor star selection
of \citet{schlaufman2014} and archival data to target nine LMC and four
SMC giants for high-resolution Magellan/MIKE spectroscopy.  These nine LMC
giants with $-2.4\lesssim[\text{Fe/H}]\lesssim-1.5$ and four SMC giants
with $-2.6\lesssim[\text{Fe/H}]\lesssim-2.0$ are the most metal-poor
stars in the Magellanic Clouds yet subject to a comprehensive abundance
analysis.  While we find that at constant metallicity these stars are
similar to Milky Way stars in their $\alpha$, light, and iron-peak
elemental abundances, both the LMC and SMC are enhanced relative to the
Milky Way in the $r$-process element europium.	These abundance offsets
are highly significant, equivalent to$3.9\sigma$ 
for the LMC, $2.7\sigma$ for the SMC, and 
$5.0\sigma$ for the complete Magellanic Cloud
sample.  We propose that the $r$-process enhancement of the Magellanic
Clouds' metal-poor stellar population is a result of the Magellanic
Clouds' isolated chemical evolution and long history of accretion from
the cosmic web combined with $r$-process nucleosynthesis on a timescale
longer than the core-collapse supernova timescale but shorter than or
comparable to the thermonuclear (i.e., Type Ia) supernova timescale.

\end{abstract}


\section{Introduction}\label{intro}

A common goal of Local Group galactic archaeology is the
use of the elemental abundances of a galaxy's metal-poor
stars to constrain its formation and early chemical evolution
\citep[e.g.,][]{beers2005,tolstoy2009,frebel2015,simon2019}.
The formation and early chemical evolution of massive spiral galaxies
like the Milky Way and M31 with $M_{\text{tot}} \sim 10^{12}~M_{\odot}$
\citep[e.g.,][]{wang2020,kafle2018} resulted from complex interactions
between inflows, star formation, stellar evolution, nucleosynthesis,
and outflows \citep[e.g.,][]{kobayashi2006,kobayashi2020}.  In the
metallicity range $-3 \lesssim [\text{Fe/H}] \lesssim -1$, stars
in both the Milky Way and M31 have supersolar abundances relative to
iron of titanium and the $\alpha$ elements magnesium, silicon, and calcium
\citep[e.g.,][]{mcwilliam1995a,mcwilliam1995b,escala2019,escala2020a,escala2020b,gilbert2020}.
The supersolar $[\alpha/\text{Fe}]$ abundance ratios observed
in these stars are thought to result from nucleosynthesis
in core-collapse supernovae that start enriching a galaxy's
interstellar medium just a few Myr after the onset of star formation
\citep[e.g,][]{woosley1995,heger2010,sukhbold2016}.  The eventual
prolific nucleosynthesis of iron-peak elements in thermonuclear (i.e.,
Type Ia) supernovae starting a few tens to 100 Myr after the onset of star
formation causes the $[\alpha/\text{Fe}]$ abundance ratio to approach
solar as $[\text{Fe/H}]$ approaches zero \citep[e.g.,][]{maoz2014}.
The $[\text{Fe/H}]$ value at which $[\alpha/\text{Fe}]$ begins to decline
(sometimes called the ``knee'') therefore corresponds to a point in time
a few tens to 100 Myr after the onset of star formation in a stellar
population \citep[e.g.,][]{tinsley1979}.  The implication is that both
the Milky Way and M31 transitioned from forming stars with $[\text{Fe/H}]
\sim -3$ to forming stars with $[\text{Fe/H}] \sim -1$ in less than
about 100 Myr.  The Milky Way and M31's high masses and star formation
rates necessary to drive chemical evolution from $[\text{Fe/H}] \sim -3$
to $[\text{Fe/H}] \sim -1$ in less than 100 Myr suggest that even rare
classes of supernovae should contribute to their chemical evolution.

\subsection{Metal-poor Star Formation in Classical and Ultra-faint Dwarf Spheriodal Satellites}

The formation and chemical evolution of the Milky Way's
satellite classical and ultra-faint dwarf spheroidal
(dSph) galaxies\footnote{From here all references to classical and 
dSph refer to satellite galaxies.} with masses enclosed inside their half-light radii
$M_{1/2}$ in the range $10^{5}~M_{\odot} \lesssim M_{1/2} \lesssim
10^{8}~M_{\odot}$ appear to have been simpler.\footnote{While
the distances to M31's dSph galaxies make them
more challenging to study, at constant mass they are generally
similar to the Milky Way's satellites \citep[e.g.,][]{wojno2020}.}
The Milky Way's classical dSph galaxies with
$10^{7}~M_{\odot} \lesssim M_{1/2} \lesssim 10^{8}~M_{\odot}$ have
extended star formation histories, broad metallicity distributions,
and no discernible downturns in $[\alpha/\text{Fe}]$ with increasing
$[\text{Fe/H}]$ in the range $-3 \lesssim [\text{Fe/H}] \lesssim -1$
\citep[e.g.,][]{tolstoy2009,kirby2011a,kirby2011b}.  While simple chemical
evolution models are unable to explain all of their observed properties,
significant inflows of unenriched gas are usually necessary.  Moreover,
the importance of unenriched gas inflow relative to gas outflows appears
to increase with galaxy luminosity.

Like the Milky Way's more massive classical dSph
galaxies, its ultra-faint dSph galaxies with
$10^{5}~M_{\odot} \lesssim M_{1/2} \lesssim 10^{7}~M_{\odot}$ have
extended star formation histories and broad metallicity distributions
\citep[e.g.,][]{brown2012,brown2014,vargas2013,weisz2014}.
In contrast to their more massive counterparts, the Milky Way's
ultra-faint dSph galaxies stopped forming stars
after only a few Gyr.  When considered as a population, stars in
the Milky Way's ultra-faint dSph galaxies show
a decline in $[\alpha/\text{Fe}]$ with increasing $[\text{Fe/H}]$
in the range $-3 \lesssim [\text{Fe/H}] \lesssim -1$ as expected if
star formation persisted for more than about 100 Myr and thermonuclear
supernovae had an opportunity to contribute to their chemical evolution
\citep[e.g.,][]{vargas2013}.  Chemical evolution models for ultra-faint
dSph galaxies have suggested that they were an order of magnitude less
efficient than classical dSph galaxies at turning their cosmological
complement of baryons into stars \citep[e.g.,][]{vincenzo2014}.
When combined with their low masses, the low star formation efficiencies
realized in ultra-faint dSph galaxies indicate that the probability of
a rare class of supernova occurring in a single ultra-faint dSph galaxy
is small \citep[e.g.,][]{ji2016a,ji2016b}.

The formation of metal-poor stars with $[\text{Fe/H}] \lesssim -2$
will persist while a galaxy's star formation rate is small compared to
the inflow rate of unenriched gas from the cosmic web.  Alternatively,
a galaxy will cease forming metal-poor stars when its star formation
rate is large relative to the inflow rate of unenriched gas from the
cosmic web.  The interplay between star formation and the accretion
of unenriched gas is therefore the key determinant of the duration of
metal-poor star formation.  The available data described above suggests
that the chemical evolution of the Milky Way and M31 advanced rapidly
through the range $-3 \lesssim [\text{Fe/H}] \lesssim -1$.  In contrast,
the chemical evolution of the Milky Way's classical and
ultra-faint dSph galaxies moved slowly through the range $-3 \lesssim
[\text{Fe/H}] \lesssim -1$.  While the physical processes that halted
the inflow of unenriched gas from the cosmic web and therefore ended
the era of metal-poor star formation are not directly observable, it is
possible to directly study the quenching of metal-rich star formation
in nearby galaxies.

\subsection{Quenching Metal-poor Star Formation}
Galaxies can be roughly categorized using their morphologies
and rest-frame optical colors into two classes: spiral
or irregular galaxies that are part of a ``blue cloud''
and elliptical galaxies that are part of a ``red sequence''
\citep[e.g.,][]{strateva2001,blanton2003,baldry2004,wuyts2011,vanderwel2014}.
While both morphologies and rest-frame optical colors are manifestations
of galaxies' star formation rates, the physical processes responsible
for quenching star formation and therefore moving galaxies from the
blue cloud to the red sequence are complex and uncertain.  It has been
shown that both galaxy mass and environment contribute to quenching star
formation \citep[e.g.,][]{peng2010}.  The implication is that processes
both intrinsic and extrinsic to individual galaxies play some role in
heating and/or removing cold gas from halos and therefore quenching
star formation.

Possible processes responsible for intrinsic
quenching are stellar winds \& supernovae explosions
\citep[e.g.,][]{webster2014,webster2015,blandhawthorn2015},
AGN feedback \citep[e.g.,][]{fabian2012}, or the shock-heating
of cold gas falling onto a galaxy from the cosmic web
\citep[e.g.,][]{birnboim2003,keres2005,dekel2006}.  The first is
expected to be important for lower-mass galaxies with shallower
gravitational potentials, while the latter two are expected to be
important for massive galaxies with deeper gravitational potentials.
Extrinsic quenching of star formation is thought to result from the
removal of a galaxy's cold gas by the reionization of the universe
\citep[e.g.,][]{bullock2000,somerville2002,benson2002a,benson2002b},
ram-pressure stripping \citep[e.g.,][]{gunn1972}, or the interdiction
of its supply of cold gas in a process called ``strangulation''
\citep[e.g.,][]{larson1980,vandenbosch2008,putman2021}.

\subsection{The Magellanic Clouds}
The Magellanic Clouds occupy a mass range between the Milky Way's
lower-mass ultra-faint and classical dSph 
galaxies and the higher-mass Milky Way and M31.  Before their
interactions with the Milky Way, the Large and Small Magellanic
Clouds had masses $M_{\text{tot}} \sim 10^{11}~M_{\odot}$
and $M_{\text{tot}} \sim 10^{10}~M_{\odot}$ respectively
\citep[e.g.,][]{vandermarel2014,erkal2019,bekki2009,diteodoro2019}.
Unlike most of the Milky Way's classical and ultra-faint dSph galaxies
that have well-defined apocenters and therefore have been inside the
Milky Way's virial radius for a few orbital periods and thus several
Gyr \citep[e.g.,][]{gaia2018,fritz2018,simon2018}, the orbits of the
LMC and SMC support the idea that they only recently entered the Milky
Way's virial radius.  This recent infall inference is especially robust
if the Magellanic Clouds have been part of a bound pair for several Gyr
\citep[e.g.,][]{besla2007,besla2010,kallivayali2013}.

The idea that the Magellanic Clouds evolved in isolation for most
of their histories is supported by their inferred star formation and
chemical evolution histories.  Models of the Magellanic Clouds' star
formation histories suggest initial surges of star formation followed
by relatively low levels of star formation until bursts a few Gyr in
the past \citep[e.g.,][]{harris2004,harris2009}.  These bursts of star
formation are plausibly associated with the Magellanic Clouds' first
interaction with the Milky Way.  Masses in excess of $M_{\text{tot}}
\sim 10^{10}~M_{\odot}$ combined with flat age--metallicity relations for
both the LMC at $[\text{Fe/H}] \approx -1.2$ between 12 and 5 Gyr in the
past and the SMC at $[\text{Fe/H}] \sim -1$ between 8 and 3 Gyr in the
past require the accretion of significant amounts of unenriched gas.
\citet{nidever2020} have suggested that $[\alpha/\text{Fe}]$ begins
to decline in the Magellanic Clouds at $[\text{Fe/H}] \lesssim -2.2$
as well, implying that the chemical evolution of both the LMC and SMC
had advanced no further than $[\text{Fe/H}] \approx -2.2$ after about
100 Myr.  While the chemical evolution of the SMC has not yet been studied
in detail, models of the LMC's chemical evolution confirm the importance
of the accretion of low-metallicity gas \citep[e.g.,][]{bekki2012}.

All of these data are consistent with a scenario in which the
Magellanic Clouds spent the majority of their histories beyond the
Milky Way's virial radius.  They were therefore able to accrete gas
from the cosmic web because of their intermediate masses and relative
isolation.  They were then able to retain that gas against stellar and
supernovae feedback because of their intermediate-depth potential wells
\citep[e.g.,][]{jahn2019,santos2021}.  Indeed, \citet{geha2012} have
shown that in the mass range of the Magellanic Clouds internal processes
appear to be unable to quench star formation.  At the same time, they
were not massive enough to be significantly affected by AGN feedback or
the shock-heating of infalling unenriched gas.  The net result is that
metal-poor star formation at rates high enough to realize even rare
supernovae likely extended for a longer time in the Magellanic Clouds
than in the Milky Way, M31, or most of their satellite dSph galaxies.

\subsection{The Origin of $r$-process Elements}
The Magellanic Clouds' high rates (relative to the Milky Way's
dSph satellite galaxies) and extended durations of metal-poor star
formation (relative to the Milky Way and M31) provide an opportunity
to identify the astrophysical origin of the rapid neutron-capture
process or $r$-process nucleosynthesis.  While some small amount of
$r$-process nucleosynthesis probably occurs in ordinary core-collapse
supernovae \citep[e.g.,][]{ji2016b,casey2017}, a rare class of
explosive event is thought to be responsible for most $r$-process
nucleosynthesis.  Evidence for a rare-but-prolific astrophysical origin
for $r$-process nucleosynthesis comes from sea floor $^{244}\text{Pu}$
\citep[e.g.,][]{wallner2015,hotokezaka2015}, early solar system
$^{244}\text{Pu}$ \& actinide abundances \citep{bartos2019},
the $r$-process abundances of stars in ultra-faint dSph galaxies
\citep[e.g.,][]{ji2016a}, and models of galactic chemical evolution
\citep[e.g.,][]{shen2015,vandevoort2015,naiman2018,macias2018}.
Two types of models have been suggested to explain the existence
of $r$-process enriched metal-poor stars: (1) unusual core-collapse
supernovae like collapsars or magnetorotationally driven supernovae
\citep[e.g.,][]{pruet2004,fryer2006,winteler2012,siegel2019}
and (2) compact object mergers involving a neutron star
\citep[e.g.,][]{lattimer1974,symbalisty1982}.  While both classes of
models produce $r$-process nucleosynthesis in rare events, in the unusual
core-collapse supernova model $r$-process nucleosynthesis should be prompt
on a timescale comparable to the core-collapse supernova timescale.
On the other hand, in the neutron star merger model $r$-process
nucleosynthesis events should occur only after some delay of uncertain
duration from the onset of star formation.

One way to evaluate both the reality and duration of a delay in the onset
of $r$-process nucleosynthesis would be to compare the relative occurrence
of $r$-process enhanced stars in both the Milky Way and the Magellanic
Clouds at $[\text{Fe/H}] \sim -2$ and $[\alpha/\text{Fe}] \approx 0.4$.
Both galaxies are massive enough and experience sufficient metal-poor
star formation for rare supernovae to occur.  The key difference is that
chemical evolution in the Milky Way moved through the range $-3 \lesssim
[\text{Fe/H}] \lesssim -1$ before thermonuclear supernovae could become
significant contributors to its chemical evolution.  In the same amount
of time, chemical evolution in the Magellanic Clouds advanced no further
than $[\text{Fe/H}] \approx -2.2$.  If the occurrence of $r$-process
enhanced metal-poor stars is the same in both the Milky Way and the
Magellanic Clouds, then $r$-process nucleosynthesis occurs on a timescale
comparable to the core-collapse supernova timescale.  If $r$-process
enhanced metal-poor stars are more common in the Magellanic Clouds, then
$r$-process nucleosynthesis occurs with a delay between the core-collapse
supernova and thermonuclear supernova delay timescales in the range of
a few to 100 Myr.

In this paper, we isolate and then study the stellar parameters and
detailed elemental abundances of the most metal-poor stars yet subject to
comprehensive abundance analyses in both the Large and Small Magellanic
Clouds.  We describe in Section \ref{sample_obs} our sample selection
and high-resolution follow-up observations.  We then infer stellar
parameters from those spectra and all available photometric data in
Section \ref{stellar_prop}.  We report each star's detailed elemental
abundances in Section \ref{chem_abund}.  We review our results and their
implications in Section \ref{discussion}.  We conclude by summarizing
our findings in Section \ref{conclusion}.

\section{Sample Selection and Observations}\label{sample_obs}

We selected our initial list of candidate metal-poor giants using
a variant of the \citet{schlaufman2014} infrared metal-poor star
selection on 2MASS near-infrared and Spitzer SAGE or AllWISE mid-infrared
photometry \citep{skrutskie2006,meixner2006,wright2010,mainzer2011}.
We required $|[3.6] - [4.5]| < 0.04$ in Spitzer SAGE or $|W1 - W2| < 0.04$
in AllWISE.  While we used a near-infrared color cut $0.45 < J-H < 0.60$
in \citet{schlaufman2014}, we extended that color cut to $0.45 < J-H <
0.70$ in the Magellanic Clouds to include metal-poor giants closer to
the tip of the red giant branch.  The only Spitzer SAGE data quality
cut we imposed was $\text{\texttt{closeflag}} = 0$ to require all
sources to have no detected sources within 3\arcsec of our candidates.
Because we were confident that we could exclude LMC and SMC variable
stars identified as such by the Optical Gravitational Lensing Experiment
(OGLE)\footnote{The OGLE survey and its variable star discoveries in the
Magellanic Clouds have been described in \citet{udalski2008,udalski2015},
\citet{soszyski2015,soszyski2016,soszyski2017,soszyski2018}, and
\citet{pawlak2016}.}, we did not use in our selection the AllWISE
\texttt{var\_flg} metadata.  We also ignored AllWISE \texttt{moon\_lev}
metadata.

The procedure described above produced an initial candidate list that
we then examined using SIMBAD.  We found literature data for five
of our candidates from \citet{cole2005} (2MASS J05224088-6951471 and
2MASS J05242202-6945073), \citet{carrera2008} (2MASS J05141665-6454310),
\citet{pompeia2008}/\citet{vanderswaelmen2013} (2MASS J05133509-7109322)
or \citet{dobbie2014} (2MASS J00263959-7122102).  We prioritized for
high-resolution follow-up our candidates with low inferred metallicities
from those sources.  We also confirmed the metal-poor nature of four of
our candidates (2MASS J00251849-7140074, 2MASS J00262394-7128549, 2MASS
J00263959-7122102, and 2MASS J00273753-7125114) using data described in
\citet{mucciarelli2014} collected with UT2 of the Very Large Telescope
and its GIRAFFE instrument fed by the Fibre Large Array Multi Element
Spectrograph \citep[FLAMES;][]{pasquini2002} in a field centered on the
ancient SMC globular cluster NGC 121.\footnote{Based on observations
collected at the European Southern Observatory under ESO program
086.D-0665(A) obtained from the ESO Science Archive Facility under request
number 284138.}  NGC 121 has heliocentric radial velocity $v_r = 138 \pm
4$ km s$^{-1}$, $[\text{Fe/H}] = -1.41 \pm 0.04$, and distance $d = 63.4
\pm 0.8$ kpc \citep[e.g.,][]{dacosta1998,johnson2004,glatt2008}.  These
four metal-poor giants have positions well beyond the cluster's tidal
radius, and as we will show have radial velocities, metallicities, and/or
distances that rule out an association with NGC 121.  We supplemented our
mid-infrared selected candidates with an additional three candidates from
\citet{carrera2008} with low \ion{Ca}{2}-inferred metallicities (2MASS
J05121686-6517147, 2MASS J05143154-6505189, and 2MASS J05150380-6647003).

\begin{figure*}
\plottwo{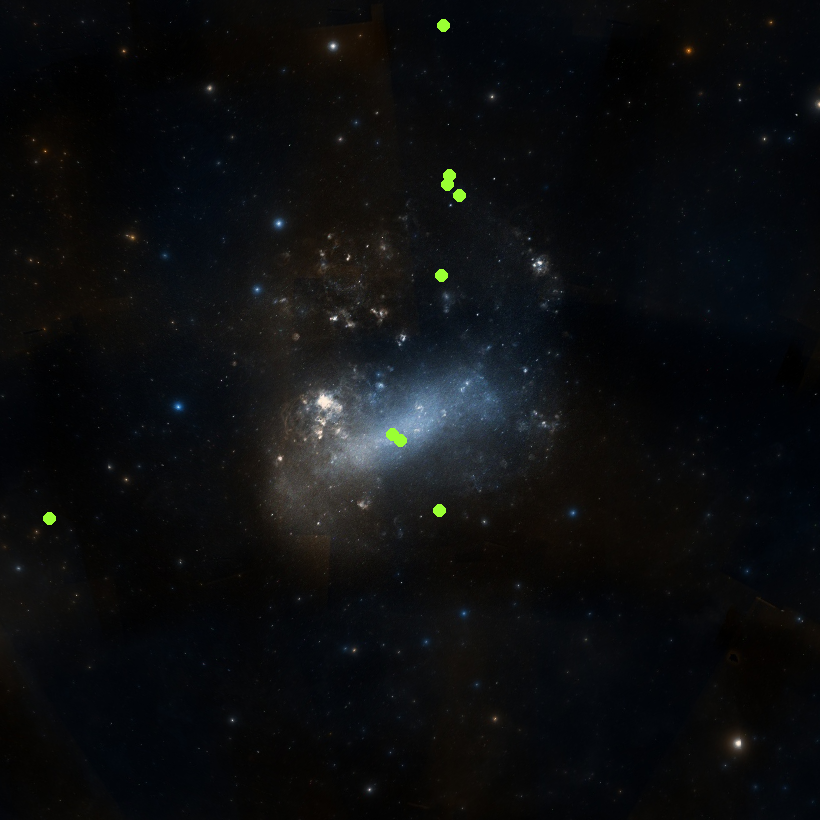}{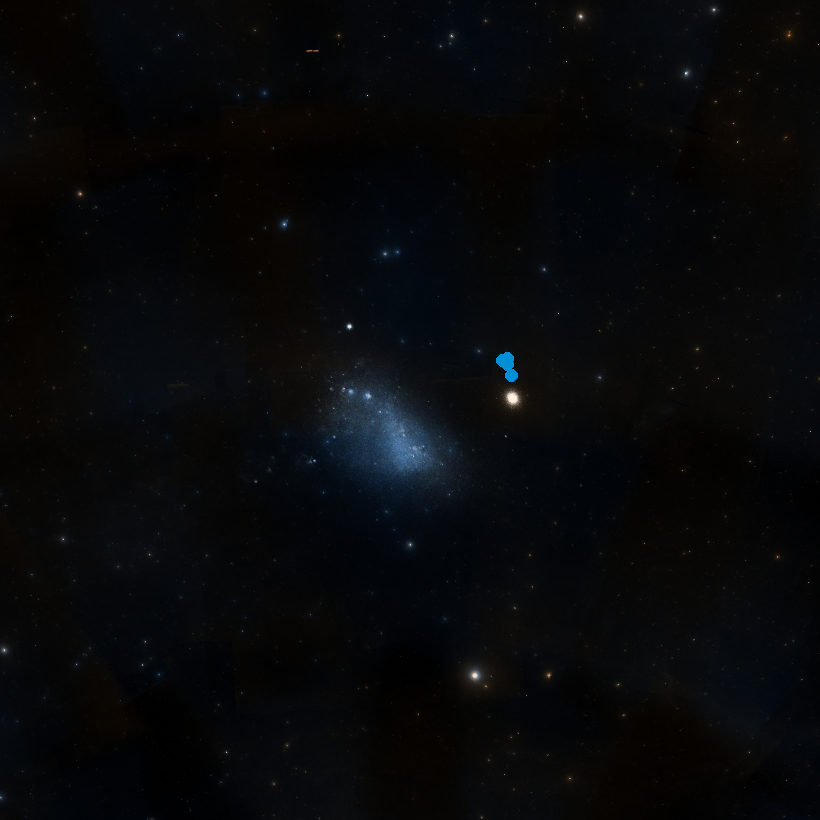}
\caption{Digitized Sky Survey images of the Magellanic Clouds.  Each image
is approximately 15$^{\circ}$ on a side with north at the top and east to
the left.  Left: Large Magellanic Cloud with the locations of the nine
metal-poor giants we followed-up with high-resolution Magellan/MIKE
spectroscopy superimposed as green points.  Right: Small Magellanic
Cloud with the locations of the four metal-poor giants we followed-up
with high-resolution Magellan/MIKE spectroscopy superimposed as blue
points.\label{on_sky}}
\end{figure*}

We subsequently found that these seven LMC and four SMC stars
all have metallicities in the range $-2.6 \lesssim [\text{Fe/H}]
\lesssim -1.7$, leaving a significant metallicity gap between
the most metal-rich LMC stars in our sample and the metal-poor
LMC stars analyzed by \citet{vanderswaelmen2013}.  To close that
gap, we selected two LMC stars from the fourth phase of the Sloan
Digital Sky Survey's extension of the Apache Point Observatory
Galactic Evolution Experiment to the southern sky \citep[APOGEE-2;
][]{blanton2017,majewski2017,zasowski2017,wilson2019}.
These stars had SDSS Data Release 16 APOGEE Stellar Parameters
and Chemical Abundances Pipeline \citep[ASPCAP; ][]{garcieperez2016, ahumada2020}
inferred metallicities in the range $-1.7 \lesssim [\text{Fe/H}]
\lesssim -1.5$ (2MASS J05160009-6207287 and 2MASS J06411906-7016314).
We plot the locations of these 13 stars on Digitized Sky Survey images
of the LMC and SMC in Figure \ref{on_sky}.

We followed up these 13 stars with the Magellan Inamori Kyocera Echelle
(MIKE) spectrograph on the Magellan Clay Telescope at Las Campanas
Observatory \citep{bernstein2003,shectman2003}.  We used either the
0\farcs7~or 1\farcs0~slits and the standard blue and red grating azimuths,
yielding spectra between 335 nm and 950 nm with resolution $R \approx
40,\!000/28,\!000$ in the blue and $ R \approx 31,\!000/22,\!000$
in the red for the 0\farcs7/1\farcs0~slits.  We collected all
calibration data (e.g., bias, quartz \& ``milky" flat field, and ThAr
lamp frames) in the afternoon before each night of observations.
We present a log of these observations in Table~\ref{obs_log}.
We reduced the raw spectra and calibration frames using the
\texttt{CarPy}\footnote{\url{http://code.obs.carnegiescience.edu/mike}}
software package \citep{kelson2000,kelson2003,kelson2014}.  We used
\texttt{iSpec}\footnote{\url{https://www.blancocuaresma.com/s/iSpec}}
\citep{blanco-cuaresma2014,blanco-cuaresma2019} to calculate radial
velocities and barycentric corrections and normalized individual orders
using \texttt{IRAF}\footnote{\url{https://iraf-community.github.io/}}
\citep{iraf1986,iraf1993}.

\begin{deluxetable*}{lcccccccccc}
\tablecaption{Log of Magellan/MIKE Observations\label{obs_log}}
\tablewidth{0pt}
\tablehead{
\colhead{2MASS} &
\colhead{Galaxy} &
\colhead{R.A.} &
\colhead{Decl.} & 
\colhead{UT Date} &
\colhead{Start} &
\colhead{Slit} &
\colhead{Exposure} &
\colhead{$v_r$}  &
\colhead{S/N} &
\colhead{S/N} \\
\colhead{Designation} &
\colhead{} &
\colhead{(h:m:s)} &
\colhead{(d:m:s)} &
\colhead{} &
\colhead{} &
\colhead{Width} &
\colhead{Time (s)} &
\colhead{(km s$^{-1}$)} &
\colhead{$4500 \ \rm{\AA}$} &
\colhead{$6500 \ \rm{\AA}$}
}
\startdata
J00251849-7140074 & SMC & 00:25:18.47 & -71:40:07.50 & 2017 Jul 03 & 07:29:38 & 1\farcs0 & \hphantom03600 & $100.5 \pm 1.1$ & $22$ & $45$\\ 
J00262394-7128549 & SMC & 00:26:23.95 & -71:28:54.94 & 2017 Jul 02 & 08:05:08 & 1\farcs0 & 14400 & $ 133.7 \pm 0.9$ & $25$ & $60$ \\
J00263959-7122102 & SMC & 00:26:39.60 & -71:22:10.26 & 2017 Jun 29 & 07:59:59 & 1\farcs0 & \hphantom03600 & $151.3 \pm 1.0$ & $15$ & $31$\\
J00273753-7125114 & SMC & 00:27:37.55 & -71:25:11.56 & 2017 Jun 28 & 09:06:17 & 1\farcs0 & \hphantom03600 & $149.4 \pm 2.1 $ & $15$ & $47$ \\
J05121686-6517147 & LMC & 05:12:16.85 & -65:17:14.76 & 2018 Jan 07 & 05:37:02 & 1\farcs0 & \hphantom03584 & $336.2 \pm 1.1$ & $14$ & $50$ \\
J05133509-7109322 & LMC & 05:13:35.10 & -71:09:32.18 & 2018 Jan 08 & 04:44:02 & 1\farcs0 & \hphantom05400 & $186.9 \pm 1.0$  & $18$ & $45$ \\
J05141665-6454310 & LMC & 05:14:16.65 & -64:54:31.30 & 2018 Jan 07 & 00:56:51 & 1\farcs0 & 19800 & $298.9 \pm 1.4$ & $30$ & $65$ \\
J05143154-6505189 & LMC & 05:14:31.55 & -65:05:19.02 & 2018 Jan 08 & 03:41:25 & 1\farcs0 & \hphantom03600 & $304.3 \pm 1.3$ & $15$ & $35$\\
J05150380-6647003 & LMC & 05:15:03.79 & -66:47:00.28 & 2018 Jan 07 & 04:03:27 & 1\farcs0 & \hphantom05400 & $356.8 \pm 1.5$ & $20$ & $60$ \\
J05160009-6207287 & LMC & 05:16:00.09 & -62:07:28.70 & 2020 Oct 21 & 07:24:32 & 1\farcs0 & \hphantom06235 & $266.0 \pm 1.1$ & $25$ & $60$\\
J05224088-6951471 & LMC & 05:22:40.89 & -69:51:47.08 & 2016 Jan 20 & 01:23:29 & 0\farcs7 & 17410 & $265.3 \pm 3.5$ & $30$ & $75$\\
J05242202-6945073 & LMC & 05:24:22.03 & -69:45:07.22 & 2016 Jan 18 & 01:16:45 & 0\farcs7 & 25200 & $309.0 \pm 1.3$ & $15$ & $50$\\
J06411906-7016314 & LMC & 06:41:19.05 & -70:16:31.42 & 2020 Oct 21 & 04:59:20 & 1\farcs0 & \hphantom07200 & $320.3 \pm 1.1$ & $25$ & $76$
\enddata
\end{deluxetable*}

\section{Stellar Properties}\label{stellar_prop}

We use both the classical excitation/ionization balance approach and
isochrones to infer photospheric stellar 
parameters for the 13 stars listed in Table~\ref{obs_log}.  Isochrones
are especially useful for effective temperature $T_{\text{eff}}$
inferences in this case, as high-quality multiwavelength photometry
from the ultraviolet to the mid infrared is available for both the LMC
and SMC.  Similarly, the known distances of the Magellanic Clouds make the
calculation of surface gravity $\log{g}$ via isochrones straightforward.
With both $T_{\text{eff}}$ and $\log{g}$ available via isochrones, 
the equivalent widths 
of iron lines can be used to self-consistently determine metallicity
$[\text{Fe/H}]$ and microturbulence $\xi$.

We follow the algorithm outlined in \citet{reggiani2020}.  Using atomic
absorption line data from \citet{ji2020} based on the \texttt{linemake}
code\footnote{\url{https://github.com/vmplacco/linemake}}
\citep{sneden2009,sneden2016} maintained by Vinicius Placco and Ian
Roederer, we first measure the equivalent widths of \ion{Fe}{2} atomic
absorption lines by fitting Gaussian profiles with the \texttt{splot}
task in \texttt{IRAF} to our continuum-normalized spectra.  We use only
\ion{Fe}{2} lines with equivalent widths smaller than 150 m\AA~in our 
photospheric stellar parameter inferences
to avoid the possibly important effects of departures from local
thermodynamic equilibrium (LTE) on \ion{Fe}{1}-based inferences.
Whenever necessary, we use the \texttt{deblend} task to disentangle
absorption lines from adjacent spectral features.  We report our input
atomic data, measured equivalent widths, and inferred abundances in
Table \ref{measured_ews}.

To evaluate the relative importance of departures from LTE on
\ion{Fe}{1}- and \ion{Fe}{2}-based metallicity inferences, we used the
grid of corrections to abundances derived under the assumptions of LTE for
departures from LTE (i.e., non-LTE corrections) from \citet{amarsi2016}
for stars with $\log{g} = 1.5$.  For all of our stars, the non-LTE
corrections to our inferred \ion{Fe}{2}-based metallicities were less than
0.1 dex and therefore smaller than the overall metallicity uncertainties
we adopted.  On the other hand, corrections for individual \ion{Fe}{1}
lines can be as large as 0.3 dex.  We therefore argue that the metallicity
inference approach we followed based on \ion{Fe}{2} lines is likely to
be more accurate and precise than one based on \ion{Fe}{1} lines.

\begin{deluxetable*}{llccccc}
\tablecaption{Atomic Data, Equivalent-width Measurements, and
Individual-line Abundance Inferences\label{measured_ews}}
\tablewidth{0pt}
\tablehead{
\colhead{2MASS} &
\colhead{Wavelength} &
\colhead{Species} &
\colhead{Excitation Potential} &
\colhead{log($gf$)} &
\colhead{EW} &
\colhead{$\log_\epsilon(\rm{X})$} \\
\colhead{Designation} &
\colhead{(\AA)} &
&
\colhead{(eV)}
&
&
(m\AA)
&}
\startdata
J05121686-6517147 & $5682.633$ & \ion{Na}{1} & $2.102$ & $-0.706$ & $15.50$ & $4.166$\\ 
J05133509-7109322 & $5682.633$ & \ion{Na}{1} & $2.102$ & $-0.706$ & $10.00$ & $4.090$\\ 
J05133509-7109322 & $5688.203$ & \ion{Na}{1} & $2.104$ & $-0.406$ & $25.20$ & $4.250$\\ 
J05133509-7109322 & $5895.924$ & \ion{Na}{1} & $0.000$ & $-0.194$ & $247.30$ & $4.287$\\ 
J05141665-6454310 & $5682.633$ & \ion{Na}{1} & $2.102$ & $-0.706$ & $10.10$ & $4.240$\\ 
J05141665-6454310 & $5688.203$ & \ion{Na}{1} & $2.104$ & $-0.406$ & $11.00$ & $3.982$\\ 
J05143154-6505189 & $5682.633$ & \ion{Na}{1} & $2.102$ & $-0.706$ & $114.60$ & $5.534$\\ 
J05143154-6505189 & $5688.203$ & \ion{Na}{1} & $2.104$ & $-0.406$ & $101.80$ & $5.081$\\ 
J05150380-6647003 & $5682.633$ & \ion{Na}{1} & $2.102$ & $-0.706$ & $143.40$ & $5.597$
\enddata
\tablecomments{This table is published in its entirety in the
machine-readable format.  A portion is shown here for guidance regarding
its form and content.}
\end{deluxetable*}

We use 1D plane-parallel $\alpha$-enhanced ATLAS9 model atmospheres
\citep{castelli2004}, the 2019 version of the \texttt{MOOG}
radiative transfer code \citep{sneden1973}, and the \texttt{q$^2$}
\texttt{MOOG} wrapper\footnote{\url{https://github.com/astroChasqui/q2}}
\citep{ramirez2014} to derive with the classical excitation/ionization
balance approach an initial set of photospheric 
stellar parameters $T_{\text{eff}}$, $\log{g}$, $[\text{Fe/H}]$,
and $\xi$.  Since photospheric stellar
parameters inferred via the classical excitation/ionization balance
approach differ from those derived using photometry and distance
information \citep[e.g.,][]{korn2003,mucciarelli2020},
we then use the \texttt{isochrones}
package\footnote{\url{https://github.com/timothydmorton/isochrones}}
\citep{morton2015} in an iterative process to self-consistently infer
photospheric and fundamental stellar parameters
for each star using as input:
\begin{enumerate}
\item
photospheric stellar parameters from the
classical excitation/ionization balance approach (on the first iteration)
or the reduced equivalent width balance approach (on all subsequent
iterations);
\item
$g$, $r$, $i$, and $z$ magnitudes and associated uncertainties from Data
Release (DR) 2 of the SkyMapper Southern Sky Survey \citep{onken19};
\item
$J$, $H$, and $K_{\text{s}}$ magnitudes and associated uncertainties
from the 2MASS Point Source Catalog \citep{skrutskie2006};
\item
$W1$ and $W2$ magnitudes and associated uncertainties from the
Wide-field Infrared Survey Explorer (WISE) AllWISE Source Catalog
\citep{wright2010,mainzer2011};
\item
reddening inferences from \citet{skowron2021};
\item
distances inferred via the surface brightness--color relation
for eclipsing binaries in the LMC \citep{pietrzynski2019} and
SMC \citep{graczyk2014}.  For the SMC, we account for a possibly
significant line-of-sight depth inferred by \citet{scowcroft2016} using
the mid-infrared period--absolute magnitude relation for Cepheids (i.e.,
the Leavitt Law).
\end{enumerate}
We use \texttt{isochrones} to fit the MESA Isochrones and Stellar Tracks
\cite[MIST;][]{dotter2016,choi2016,paxton2011,paxton2013,paxton2015,paxton2018,paxton2019}
library to these data using
\texttt{MultiNest}\footnote{\url{https://ccpforge.cse.rl.ac.uk/gf/project/multinest/}}
\citep{feroz2008,feroz2009,feroz2019}.  We restrict the MIST library
to stellar ages $\tau$ in the range 8.0 Gyr $\leq \tau \leq$ 13.721
Gyr and distances $d$ to the 3-$\sigma$ ranges including both
statistical and systematic uncertainties for the LMC and SMC from
\citet{pietrzynski2019} and \citet{graczyk2014} respectively accounting
for the possibly significant line-of-sight depth for the SMC advocated
by \citet{scowcroft2016}.  We also restrict our analysis to
extinctions $A_{V}$ in the ranges proposed by \citet{skowron2021} at
the location of each candidate.  While the metal-poor stars in
our sample generally have supersolar $\alpha$-element abundances, the
current version of the MIST isochrones were calculated using solar-scaled
abundances.  While this potential inconsistency can affect fundamental
stellar parameter inferences like mass and age, the uncertainties due
to this effect are usually smaller than the uncertainties resulting from
other sources \citep[e.g.,][]{grunblatt2021}.

For the first \texttt{isochrones} calculation, we use in the likelihood
the initial set of photospheric stellar
parameters inferred from the classical excitation/ionization balance
analysis of each spectrum along with the photometric, extinction, and
distance information described above.  We then use \texttt{isochrones}
to calculate a new set of photospheric and
fundamental stellar parameters that is both self-consistent and physically
consistent with stellar evolution.  We next impose the $T_{\text{eff}}$
and $\log{g}$ inferred in this way on an iteration of the reduced
equivalent width balance approach to derive $[\text{Fe/H}]$ and $\xi$
values consistent with our measured \ion{Fe}{2} equivalent widths
and our \texttt{isochrones} inferred $T_{\text{eff}}$ \& $\log{g}$.
We then execute another \texttt{isochrones} calculation, this time
using this updated set of photospheric
stellar parameters in the likelihood.  We iterate this process until the
metallicities inferred from both the \texttt{isochrones} analysis and
the reduced equivalent width balance approach are consistent within
their uncertainties (typically a few iterations).  For our final
photospheric stellar parameters, we adopt
$T_{\text{eff}}$ and $\log{g}$ from the final \texttt{isochrones}
iteration.  We infer $A_{V} \approx 1.6$ mag for the two stars
coincident with the bar of the LMC 2MASS J05224088-6951471 and 2MASS
J05242202-6945073, possibly suggesting that they are on the far side of
the LMC.

We prefer stellar parameters inferred using both isochrones and
spectroscopy to stellar parameters derived from spectroscopy alone.
Isochrone analyses can make use of constraints from multiwavelength
photometry as well as distance and extinction information impossible
to leverage with the classic spectroscopic method.  All of these data
are available for the stars in our sample.  Unlike $T_{\text{eff}}$
and $\log{g}$ values derived from spectroscopy alone, $T_{\text{eff}}$
and $\log{g}$ values deduced from isochrone analyses are guaranteed
to be self-consistent and in principle realizable during stellar
evolution (to the degree of accuracy achievable in 
by the stellar models).  It has long been known that photospheric stellar
parameters inferred for metal-poor giants using the classical approach
differ from those derived using photometry and parallax information
\citep[e.g.,][]{korn2003,frebel2013,mucciarelli2020}.  Since LTE is
almost always assumed in the calculation of the model atmospheres used
to interpret equivalent width measurements, these differences are often
attributed to the violation of the assumptions of LTE in the photospheres
of metal-poor giants.

We derive the uncertainties in our adopted $[\text{Fe/H}]$ and $\xi$
values due to the uncertainties in our adopted $T_{\text{eff}}$
and $\log{g}$ values using a Monte Carlo simulation.  We randomly
sample self-consistent pairs of $T_{\text{eff}}$ and $\log{g}$ from our
\texttt{isochrones} posteriors and calculate the values of $[\text{Fe/H}]$
and $\xi$ that produce the best reduced equivalent width balance given
our \ion{Fe}{2} equivalent width measurements.  We save the result of
each iteration and find that the contributions of $T_{\text{eff}}$
and $\log{g}$ uncertainties to our final $[\text{Fe/H}]$ and $\xi$
uncertainties are small: about 0.01 dex for both the LMC and SMC.
We derive our adopted metallicities and uncertainties using a final
reduced equivalent width balance calculation imposing the adopted
$T_{\text{eff}}$ and $\log{g}$ values described above.  The result
of this calculation agrees well with the result of the Monte Carlo
simulation.  We report 
our adopted photospheric stellar parameters both by themselves in Table
\ref{stellar_params_simple} and together with all of our derived stellar
parameters in the comprehensive Table \ref{stellar_params}.

We define the line-by-line dispersion of our inferred iron abundances
as the standard deviation of the \ion{Fe}{2} abundances inferred from
individual \ion{Fe}{2} lines.  The uncertainties in our metallicity
inferences are dominated by this line-by-line dispersion.  We provide in
Tables \ref{stellar_params_simple} and \ref{stellar_params} our adopted
\ion{Fe}{2}-based metallicities with uncertainties derived by adding
in quadrature the relatively large uncertainties from our line-by-line
abundance dispersions and the relatively small uncertainties that can
be attributed to $T_{\text{eff}}$ and $\log{g}$ uncertainties.  We also
present in Table \ref{stellar_params} a detailed view of our adopted
photospheric and fundamental stellar parameters
for all 13 stars in our LMC and SMC samples.  All of the uncertainties
quoted in Table \ref{stellar_params} include random uncertainties only.
That this, they are uncertainties derived under the unlikely assumption
that the MIST isochrone grid we use in our analyses perfectly reproduces
all stellar properties. 

\begin{deluxetable*}{lCCCC}
\tablecaption{Adopted Photospheric Stellar Parameters}\label{stellar_params_simple}
\tablewidth{0pt}
\tablehead{
\colhead{2MASS} & \colhead{$T_{\text{eff}}$} & \colhead{$\log{g}$} & \colhead{$[\text{Fe/H}]$}
& \colhead{$\xi$} \\
\colhead{Designation} & \colhead{(K)} & \colhead{} & \colhead{}
& \colhead{(km s$^{-1}$)}}
\startdata
J00251849-7140074 & $4372^{+5}_{-3}$   & $0.65\pm0.01$ & $-2.33\pm0.30$ & $2.71\pm0.50$ \\ 
J00262394-7128549 & $4388^{+8}_{-6}$   & $0.87\pm0.01$ & $-2.02\pm0.32$ & $2.51\pm0.80$ \\ 
J00263959-7122102 & $4481^{+2}_{-1}$   & $0.73\pm0.01$ & $-2.60\pm0.39$ & $3.01\pm0.45$ \\ 
J00273753-7125114 & $4326\pm10$        & $0.61\pm0.01$ & $-2.27\pm0.26$ & $3.83\pm0.95$ \\ 
J05121686-6517147 & $4247\pm1$         & $0.55\pm0.01$ & $-2.32\pm0.36$ & $2.91\pm0.50$ \\ 
J05133509-7109322 & $4520^{+26}_{-18}$ & $1.06\pm0.02$ & $-2.15\pm0.38$ & $2.55\pm0.75$ \\ 
J05141665-6454310 & $4761^{+18}_{-15}$ & $1.43\pm0.02$ & $-2.42\pm0.23$ & $2.45\pm0.62$ \\
J05143154-6505189 & $4361^{+2}_{-1}$   & $1.06\pm0.01$ & $-1.65\pm0.35$ & $2.74\pm0.83$ \\
J05150380-6647003 & $4139^{+14}_{-13}$ & $0.64\pm0.02$ & $-1.83\pm0.46$ & $3.31\pm0.91$ \\
J05160009-6207287 & $4151\pm17$        & $0.60\pm0.02$ & $-1.62\pm0.24$ & $3.25\pm0.54$ \\
J05224088-6951471 & $4619^{+45}_{-23}$ & $1.24\pm0.03$ & $-2.14\pm0.32$ & $3.16\pm0.45$ \\
J05242202-6945073 & $4903^{+31}_{-49}$ & $1.61\pm0.03$ & $-1.99\pm0.47$ & $2.38\pm0.78$ \\
J06411906-7016314 & $4093^{+8}_{-7}$   & $0.53\pm0.01$ & $-1.54\pm0.14$ & $3.28\pm0.47$
\enddata
\tablecomments{We give our adopted photospheric stellar parameters both
by themselves here and in our comprehensive table of stellar parameters
in Table \ref{stellar_params}.}
\end{deluxetable*}

To roughly quantify the magnitude of the possible photospheric
stellar parameter systematic uncertainties resulting from our analysis,
we compared our photospheric stellar parameters with those inferred
by APOGEE's ASPCAP for the two stars 2MASS J05160009-6207287 and 2MASS
J06411906-7016314 in our sample selected from SDSS DR16 \citep{ahumada2020}.  
Our photospheric
stellar parameters are consistent with those produced by ASPCAP, as the
differences in $T_{\text{eff}}$, $\log{g}$, and $[\text{Fe/H}]$ are 15
K, 0.1 dex, and 0.01 dex, fully consistent within the ASPCAP 
reported uncertainties ($\pm100$ K, $\pm0.1$ dex, and $\pm0.03$ dex).  
The excellent agreement between these two
analyses suggests that any systematic uncertainties in our photospheric
stellar parameters resulting from our analysis procedure are small.

\section{Elemental Abundances}\label{chem_abund}

To infer the elemental abundances of several $\alpha$, light odd-$Z$,
iron-peak, and neutron-capture elements, we first measure the equivalent
widths of atomic absorption lines of \ion{Na}{1}, \ion{Mg}{1},
\ion{Si}{1}, \ion{Ca}{1}, \ion{Sc}{2}, \ion{Ti}{1}, \ion{Ti}{2},
\ion{Cr}{1}, \ion{Mn}{1}, \ion{Co}{1}, \ion{Ni}{1}, \ion{Y}{2},
\ion{Ba}{2}, \ion{La}{2}, \ion{Ce}{2}, \ion{Nd}{2}, \ion{Sm}{2}, and
\ion{Gd}{2} in our continuum-normalized spectra by fitting Gaussian
profiles with the \texttt{splot} task in \texttt{IRAF}.  We use the
\texttt{deblend} task to disentangle absorption lines from adjacent
spectral features whenever necessary.  We measure an equivalent width for
every absorption line in our line list that could be recognized, taking
into consideration the quality of a spectrum in the vicinity of a line
and the availability of alternative transitions of the same species.  If
possible we avoid lines bluer than 4500 \AA, as the signal-to-noise ratios
(S/N) in our spectra blueward of 4500 \AA~are poor.  Blueward of
5000 \AA~in MIKE's blue arm we were able to measure equivalent widths
of lines with equivalent widths greater than 10 m\AA.  Redward of 5000
\AA~in MIKE's red arm, we were able to measure equivalent widths of lines
with equivalent widths greater than 5 m\AA. These limiting S/N ratios 
applied even to our lower S/N stars. In a few cases we measured lines with 
smaller EWs for species without other 
available lines. Whenever possible, we avoid
saturated lines like \ion{Mg}{1} $b$ as well.  We consider hyperfine
splitting in our abundance inferences for \ion{Mn}{1}, \ion{Co}{1},
and \ion{Y}{2}.  We account for isotopic splitting in barium using data
from \citet{mcwilliam1998} for the three \ion{Ba}{2} lines at $5853$,
$6141$, and $6496$ \AA~that we use for our barium abundance inferences.
We use the 1D plane-parallel $\alpha$-enhanced ATLAS9 model atmospheres
and the 2019 version of \texttt{MOOG} to infer elemental abundances
based on each equivalent width measurement.  We use spectral synthesis
to infer the abundance of \ion{Eu}{2}, mostly from its 6645 \AA~line.
Although this line is not usually strong, it is clearly measurable in
10 of the 13 giants in our sample.  For the most europium-poor star in
our sample 2MASS J05141665-6454310, we used the  \ion{Eu}{2} line at
4129 \AA~for our abundance inference.

We report our input atomic data, measured equivalent widths, and
the abundances and associated uncertainties we derive in Table
\ref{measured_ews}.  We present in Table \ref{chem_abundances} our
adopted mean elemental abundances and associated uncertainties defined
as the quadrature sum of the line-by-line abundance dispersions and the
uncertainties that result from our photospheric
stellar parameter uncertainties.  We indicate typical uncertainties in
each panel of Figures \ref{alphas_fig} to \ref{neutron_capture_fig}.  Our
iterative use of both isochrones and the classical excitation/ionization
balance approach to stellar parameter inference produced very precise
stellar parameters, and that precision is reflected in the abundance
uncertainties we report in Table~\ref{chem_abundances}.

As we described in Section \ref{stellar_prop}, we have not accounted
for the possibility of systematic uncertainties in our stellar parameter
inferences.  To quantify the effect of possible systematic uncertainties
in our stellar parameter inferences on our abundance inferences, we
recalculated our abundances assuming $T_{\text{eff}}$ and $\log{g}$
uncertainties $\Delta T_{\text{eff}} = 100$ K and $\Delta \log{g}
= 0.1$ dex.  However, the effect of these inflated stellar parameter
uncertainties is small because our abundance uncertainties are dominated
by our metallicity uncertainties.

To present a more complete view of the Magellanic Clouds' metal-poor
stellar populations, we also plot in Figures \ref{alphas_fig} to
\ref{neutron_capture_fig} the abundances of stars in LMC inner disk/bar
fields \citep{pompeia2008,vanderswaelmen2013} as well as the abundances
of stars in both the LMC and SMC observed by SDSS-IV/APOGEE-2 made
available as part of SDSS DR 16 \citep[e.g.,][]{nidever2020}.  For the
latter sample, to focus on LMC and SMC members we use the heliocentric
radial velocity and proper motion cuts proposed by \citet{nidever2020}
using the APOGEE spectra themselves and Gaia DR2 astrometry
\citep{gaia2016,gaia2018,salgado2017,arenou2018,lindegren2018,luri2018}.\footnote{We
note that we did not repeat the careful membership selection described
in \citet{nidever2020}, as in this case we only use the SDSS DR16
Magellanic Clouds sample to compare with the more metal-poor stars
in our sample.}  We do not segregate the SDSS DR16 Magellanic Clouds
sample into individual LMC/SMC samples, as \citet{nidever2020} suggested
that both galaxies have similarly delayed chemical evolution histories
(especially at metallicities below $[\text{Fe/H}] \sim -2$).

\subsection{$\alpha$ Elements}\label{alpha_section}

Magnesium, silicon, calcium, and titanium are often referred to as
$\alpha$ elements.  While magnesium and calcium are indeed synthesized by
the fusion of existing nuclei and $\alpha$ particles, silicon is formed
during both hydrostatic and explosive oxygen burning.  Likewise, titanium
is a product of both explosive oxygen and silicon burning.  Silicon
and titanium are still usually included as $\alpha$ elements though,
because the chemical evolution of $[\text{Si/Fe}]$ and $[\text{Ti/Fe}]$
as functions of $[\text{Fe/H}]$ are similar to that of magnesium and
calcium \citep[e.g.,][]{nomoto2006,clayton2007,wongwathanarat2017}.

\begin{figure}
\plotone{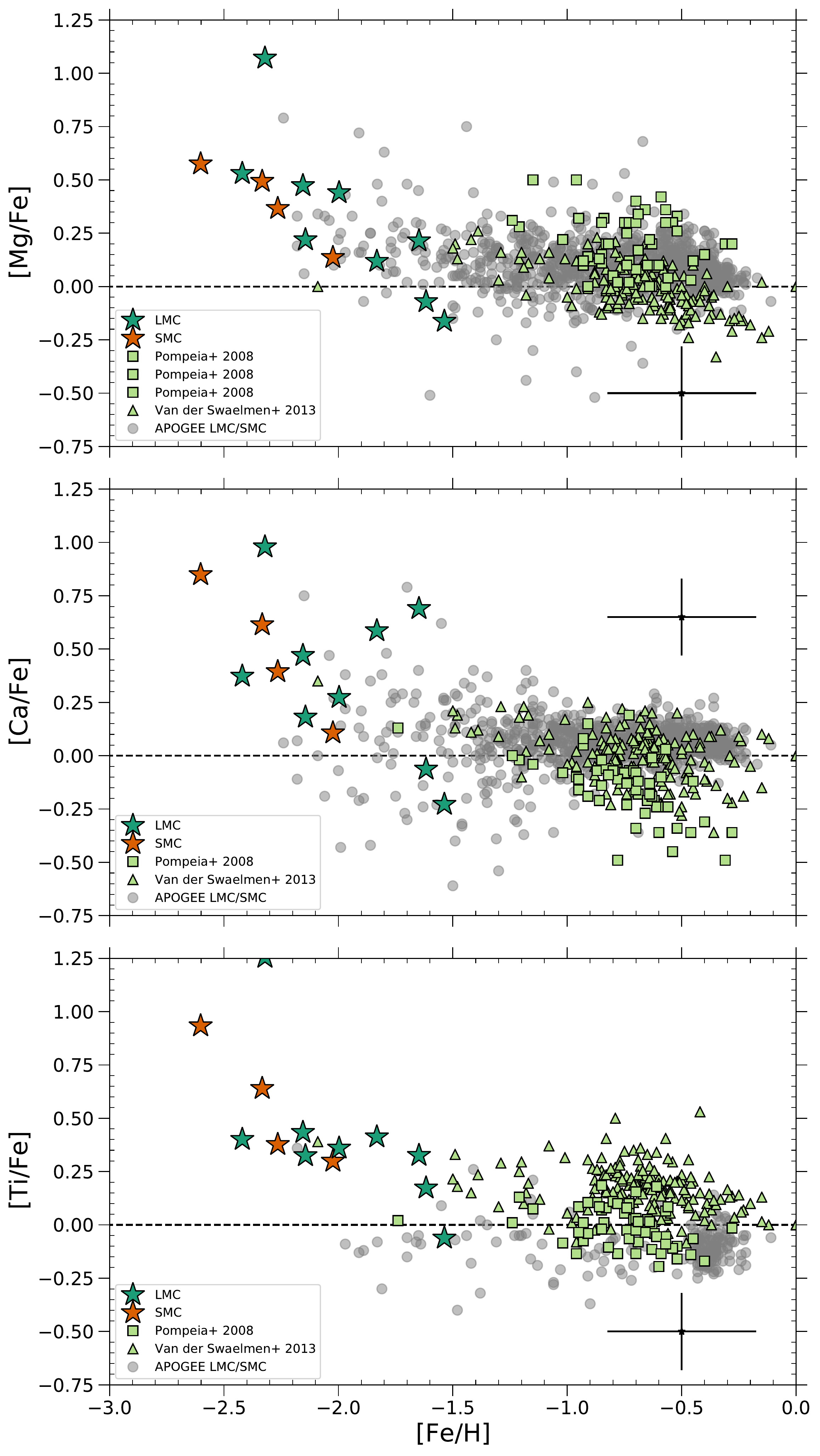}
\caption{Abundances of titanium and the $\alpha$ elements magnesium
and calcium relative to iron.  We plot as dark green stars our nine LMC
giants and as orange stars our four SMC giants.  We plot as light green
squares LMC inner disk giants from \citet{pompeia2008} and as light green
triangles LMC bar giants from \citet{vanderswaelmen2013}.  We plot as
gray circles LMC and SMC giants observed by SDSS-IV/APOGEE-2 that are
part of SDSS DR16 \citep[e.g.,][]{nidever2020}.  In the bottom-left
titanium panel, for the SDSS DR16 comparison sample we plot as points
average \ion{Ti}{1} and \ion{Ti}{2} values only for stars with both
\ion{Ti}{1} and \ion{Ti}{2} inferences. 
The point with error bars in a corner of each panel corresponds to
the averaged uncertainty of our complete 13 star sample.  Like stars
in the Milky Way and M31, we find that in the range $-2.6 \lesssim
[\text{Fe/H}] \lesssim -2.0$ LMC and SMC giants both have supersolar
$[\alpha/\text{Fe}]$ abundance ratios indicative of formation in an
environment in which thermonuclear supernovae had not yet contributed
significantly to chemical evolution.\label{alphas_fig}}
\end{figure}

We plot in Figure \ref{alphas_fig} our inferred LTE $[\alpha/\text{Fe}]$
abundances as a function of $[\text{Fe/H}]$.  We find that our inferred
magnesium, calcium, and titanium abundances are consistent with the
$\alpha$-enhanced abundances observed in metal-poor Milky Way stars.
This is true both for our LMC and SMC samples.  As we argue below,
among the $\alpha$ elements we studied our magnesium abundance inferences
are likely to be least effected by departures from the assumptions of LTE.
In the \ion{Fe}{2}-based metallicity range spanned by our Magellanic
Clouds sample $-2.6 \lesssim [\text{Fe/H}] \lesssim -1.5$, the 16th,
50th, and 84th percentiles of our $[\text{Mg/Fe}]$ distribution are
$[\text{Mg/Fe}]$ = 0.10, 0.37, and 0.53.  For the 113 Milky Way halo
stars from \citet{roederer2014} in the same \ion{Fe}{2}-based metallicity
range, the 16th, 50th, and 84th percentiles of the $[\text{Mg/Fe}]$
distribution are $[\text{Mg/Fe}]$ = 0.10, 0.26, and 0.41.  In other words,
the 1-$\sigma$ ranges of the two distributions are consistent.  While our
sample size is too small to apply the two-sided Kolmogorov-Smirnov
test often used to assess whether or not there is reason to believe
that two samples were produced by the same parent distribution, we
nevertheless performed the calculation and found a $p$-value = 0.895.
This result gives us no reason to believe that the Magellanic Clouds
$[\text{Mg/Fe}]$ distribution is distinct from that of the Milky Way
in the same metallicity range.  Applying these same steps to calcium
and titanium produce the same result and give us no reason to believe
that their distributions are distinct from that of the Milky Way in the
same metallicity range. The high abundances of magnesium and calcium we
infer are also in agreement with the increase in $[\alpha/\text{Fe}]$
with decreasing $[\text{Fe/H}]$ observed in the Magellanic Clouds
comparison sample at higher metallicities.

For the two stars in common between our sample and SDSS2 DR16,
we compared our $\alpha$-element abundances with those produced by
ASPCAP with no bad data or warning flags.  For $[\text{Mg/Fe}]$ and
$[\text{Ca/Fe}]$, both sets of abundances are consistent within the
uncertainties.  For $[\text{Si/Fe}]$, our abundances are higher than
the ASPCAP abundances.

Our inferred magnesium and calcium abundances are unlikely to
be significantly affected by departures from the assumptions of LTE.
Non-LTE corrections for LTE magnesium abundances from the INSPECT
project\footnote{\url{http://inspect-stars.com/}} based on the results
of \citet{osorio2016} are about $-0.03$ dex for the magnesium lines we
used and the photospheric stellar parameters of the stars in our sample
These corrections are too small to affect any of our conclusions based
on magnesium abundances.  Non-LTE corrections for LTE silicon abundances
$[\text{Si/Fe}] \lesssim 0.8$ based on the \citet{amarsi2017} correction
grid are about $-0.02$ dex.  For stars with abundances $[\text{Si/Fe}]
\gtrsim 0.8$ dex, non-LTE corrections can be as large as $-0.2$ dex.
Correcting for departures from LTE would therefore decrease the dispersion
in $[\text{Si/Fe}]$ for our sample.  Non-LTE corrections for LTE calcium
abundances based on the \citet{amarsi2020} correction grid are about
$-0.1$ for metal-poor giants.  Non-LTE corrections for titanium can
be as large as $+0.2$ dex \citep[e.g.,][]{bergemann2011}, but grids of
corrections for titanium are not currently available.  Overall, we argue
that magnesium is the $\alpha$ element in our sample least effected by
possible departures the assumptions of LTE.

\subsection{Light Odd-$Z$ and Iron-peak
Elements}\label{light_odd_iron_peak_section}

\begin{figure*}
\plotone{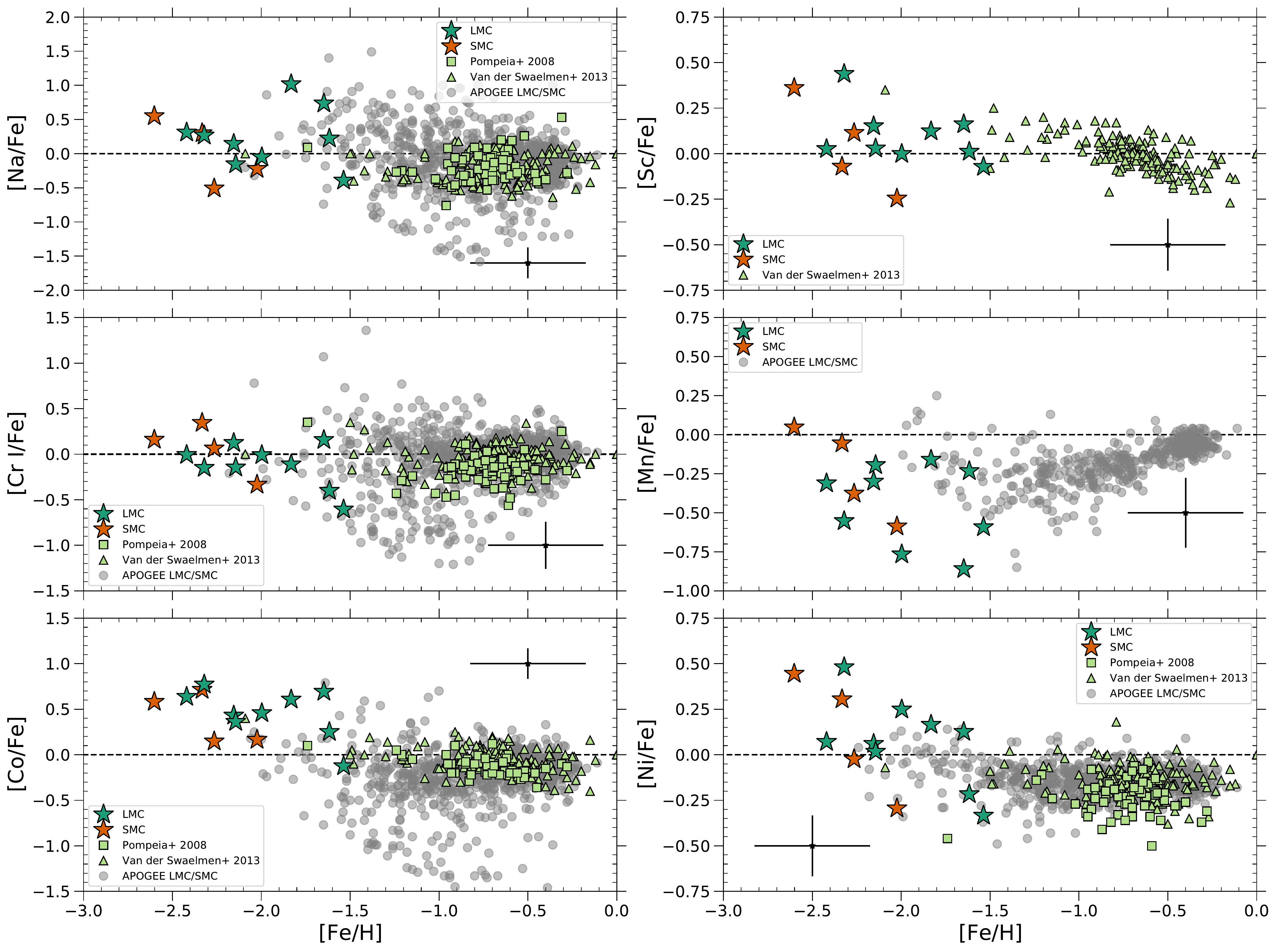}
\caption{Abundances of light odd-$Z$ elements sodium and scandium
plus iron-peak elements chromium, manganese, cobalt, and nickel.
We plot as dark green stars our nine LMC giants and as orange stars
our four SMC giants.  We plot as light green squares LMC inner disk
giants from \citet{pompeia2008} and as light green triangles LMC bar
giants from \citet{vanderswaelmen2013}.  We plot as gray circles LMC
and SMC giants observed by SDSS-IV/APOGEE-2 that are part of SDSS DR16
\citep[e.g.,][]{nidever2020}.  The point with error bars in a corner of
each panel corresponds to the mean uncertainty of our complete 13 star
sample.  Aside from elevated cobalt abundances that can be attributed to
enrichment from hypernovae, the light odd-$Z$ and iron-peak abundances
we infer are as expected for metal-poor giants in the Milky Way and its
dSph galaxies.\label{light_odd_iron_peak_fig}}
\end{figure*}

While sodium is mostly produced in core-collapse supernovae, sodium
yields are dependent on metallicity and consequently its chemical
evolution is not as easily interpreted as the chemical evolution of the
$\alpha$ elements.  Both the exact nucleosynthetic origin and chemical
evolution of scandium and potassium are likewise hard to identify and
interpret \citep[e.g.,][]{clayton2007}.  We plot in the top panels of
Figure \ref{light_odd_iron_peak_fig} our measured $[\text{Na/Fe}]$
and $[\text{Sc/Fe}]$ abundances as a function of $[\text{Fe/H}]$.
The dispersions in our inferred $[\text{Na/Fe}]$ abundances are
considerably higher than the $\alpha$ element dispersion depicted in
Figure \ref{alphas_fig}.  This is a common occurrence in studies of
metal-poor stars and is also observed at higher metallicities in the
\citet{pompeia2008}, \citet{vanderswaelmen2013}, and SDSS DR16 comparison
samples.  Our scandium abundance inferences have a dispersion comparable
to the $\alpha$ element dispersion depicted in Figure \ref{alphas_fig}.
For the two stars in common between our sample and SDSS DR16, we
compared our light odd-$Z$ abundances with those produced by ASPCAP with
no bad data or warning flags.  Both sets of abundances are consistent
within the uncertainties for $[\text{K/Fe}]$, while for $[\text{Na/Fe}]$
the ASPCAP abundances are slightly larger than our abundances.

While sodium and potassium abundance inferences can be strongly
affected by departures from LTE, scandium abundances from
\ion{Sc}{2} lines are not strongly affected by departures from LTE
\citep[e.g.,][]{zhao2016}.  Because our comparison samples did not
correct for departures from LTE or derive potassium abundances, we plot in
Figure \ref{light_odd_iron_peak_fig} LTE sodium abundances and do not plot
our potassium abundances.  In Table \ref{chem_abundances} we report our
sodium and potassium abundances inferred both under the assumptions of LTE
and corrected for the departures from LTE.  We sourced sodium abundance
corrections based on \citet{lind2011} through the INSPECT project.
We obtained potassium abundance corrections via a linear interpolation
of the \citet{reggiani2019} grid of corrections for abundances inferred
from the equivalent width of the \ion{K}{1} line at 7698 \AA.

Iron-peak elements are primarily synthesized in thermonuclear
supernovae.  The chemical evolution of chromium and nickel trace
the evolution of iron, so the nucleosynthetic origins of these
elements are thought to be similar.  We plot in the bottom panels of
Figure \ref{light_odd_iron_peak_fig} our inferred $[\text{Cr/Fe}]$,
$[\text{Mn/Fe}]$, $[\text{Co/Fe}]$, and $[\text{Ni/Fe}]$ abundances
as a function of $[\text{Fe/H}]$.  In metal-poor Milky Way stars,
both $[\text{Cr/Fe}]$ and $[\text{Ni/Fe}]$ are often observed in
solar abundance ratios.  In dSph galaxies though, significantly
non-solar $[\text{Cr/Fe}]$ and $[\text{Ni/Fe}]$ are sometimes found
\citep[e.g.,][]{ji2016b}.  While there is some evidence for supersolar
$[\text{Ni/Fe}]$ abundances and significant $[\text{Mn/Fe}]$ scatter in
our metal-poor Magellanic Cloud giants, the possible offsets are small
relative to our abundance uncertainties.  The cobalt abundances of the
metal-poor giants in our Magellanic Clouds sample  are considerably higher
than those of the comparison samples.  This is not surprising, as it is
known that there is a sharp increase in cobalt abundances with decreasing
metallicity \citep[e.g.,][]{cayrel2004,reggiani2017,reggiani2020}.
Because cobalt nucleosynthesis is dependent on core-collapse
supernovae explosion energy, the abundance increase with
decreasing metallicity could be due to enrichment by hypernovae
\citep[e.g.][]{clayton2007,kobayashi2006,kobayashi2020}.  However,
this commonly observed feature is still not reproduced by many recent
chemical evolution models \citep[e.g.,][]{prantzos2018,kobayashi2020}.
Zinc abundance inferences for these stars might help isolate the origin of
this enhancement.  For the two stars in common between our sample
and SDSS DR16, we compared our iron-peak abundances with those produced
by ASPCAP with no bad data or warning flags.  For $[\text{Cr/Fe}]$
and $[\text{Ni/Fe}]$, both sets of abundances are consistent within
the uncertainties.

Our inferred iron-peak abundances are unlikely to be significantly
affected by departures from the assumptions of LTE.  While chromium
abundances can be strongly affected by departures from the assumptions
of LTE with non-LTE corrections as large as +0.5 dex for metal-poor
giants \citep[e.g.,][]{bergemann2010}, our chromium abundances are
consistent with our comparison samples.  Non-LTE manganese corrections
for metal-poor giants are about +0.2 dex \citep[e.g.,][]{amarsi2020},
comparable to our inferred manganese uncertainties.

\subsection{Neutron-capture Elements}\label{neutron_capt_section}

Elements with atomic numbers beyond zinc are mostly produced by
neutron-capture processes either ``slow'' or ``rapid'' relative to
$\beta$ decay timescales.  The relative contributions of these $s$-
and $r$-processes to the nucleosynthesis of each neutron-capture
element are functions of metallicity.  Some elements like yttrium
and barium\footnote{Even though barium is usually used as a
tracer of the $s$-process, there can be important contributions
from $r$-process nucleosynthesis at lower metallicities
\citep[e.g.,][]{casey2017,mashonkina2019}.} are commonly used
as tracers of the $s$-process, while europium is the most easily
observed element thought to be produced exclusively in the $r$-process
\citep[e.g.,][]{cescutti2006,jacobson2013,ji2016a}.

\begin{figure*} 
\plotone{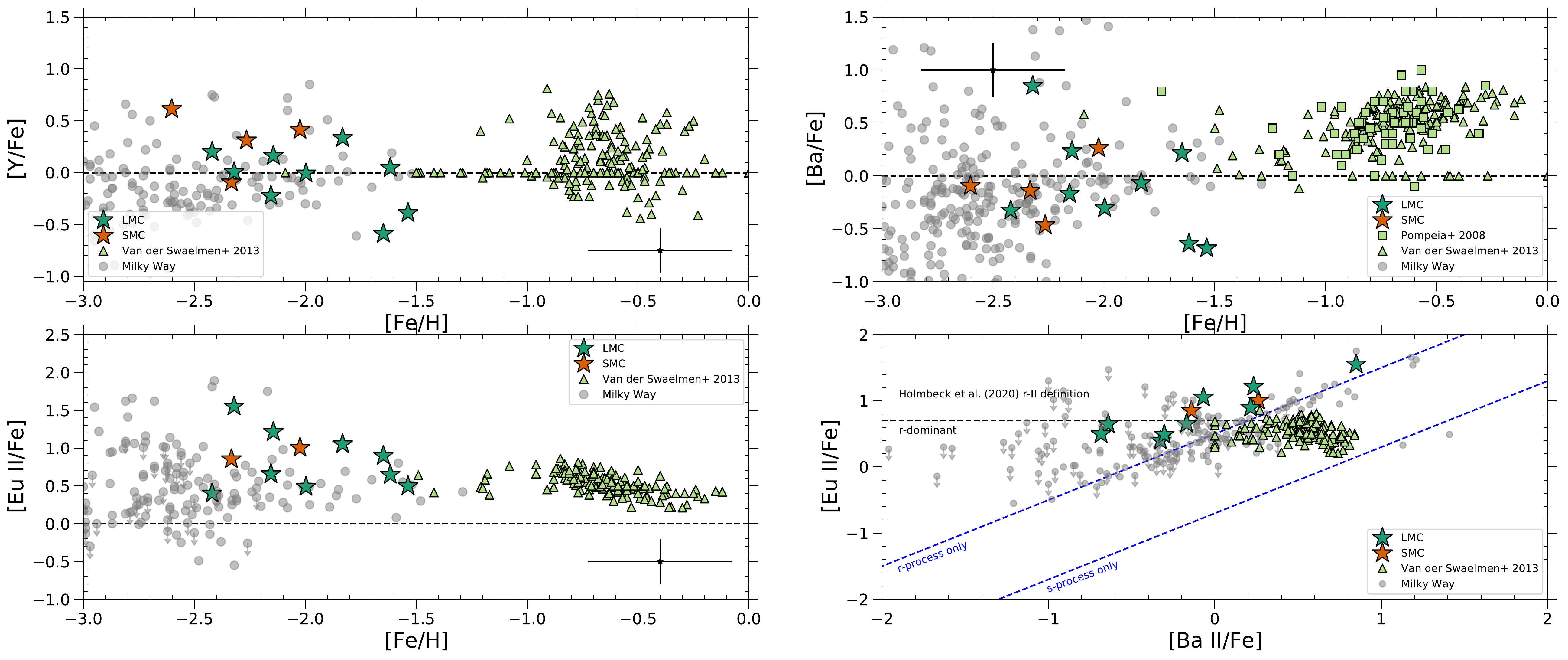}
\caption{Abundances of neutron-capture elements yttrium, barium,
and europium.  We plot as dark green stars our nine LMC giants and
as blue stars our four SMC giants.  We plot as light green squares
LMC inner disk giants from \citet{pompeia2008} and as light green
triangles LMC bar giants from \citet{vanderswaelmen2013}.  We plot as
gray circles metal-poor Milky Way giants from \citet{barklem2005} and
\citet{jacobson2015} selected for high-resolution follow-up observations
without regard to their $r$-process abundances.  The point with error
bars in a corner of each panel corresponds to the mean uncertainty
of our complete 13 star sample.  In the metallicity range $-2.6
\lesssim [\text{Fe/H}] \lesssim -1.5$, relative to the Milky Way the
$[\text{Eu/Fe}]$ abundances of the giants stars in our Magellanic Clouds
sample are uniformly high with an increased occurrence of r-II stars
with $[\text{Eu/Fe}] \ge 1.0$ and $[\text{Eu/Ba}] \geq 0$.  The stars
consistently have $[\text{Eu/Ba}]$ abundances indicative of $r$-process
nucleosynthesis.\label{neutron_capture_fig}}
\end{figure*}

We plot in Figure \ref{neutron_capture_fig} our inferred
neutron-capture element abundances as a function of $[\text{Fe/H}]$.
The abundances of the $s$-process elements yttrium and barium
inferred from \ion{Y}{2} and \ion{Ba}{2} lines are normal.
$[\text{Y/Fe}]$ abundances clustered near the solar value are
commonly observed in metal-poor Milky Way giants.  Likewise, subsolar
$[\text{Ba/Fe}]$ abundances are not uncommon and are frequently
observed in metal-poor Milky Way stars in both the halo and bulge
\citep[e.g.,][]{reggiani2017,reggiani2020,prantzos2018,kobayashi2020}.

\begin{figure*}
\plotone{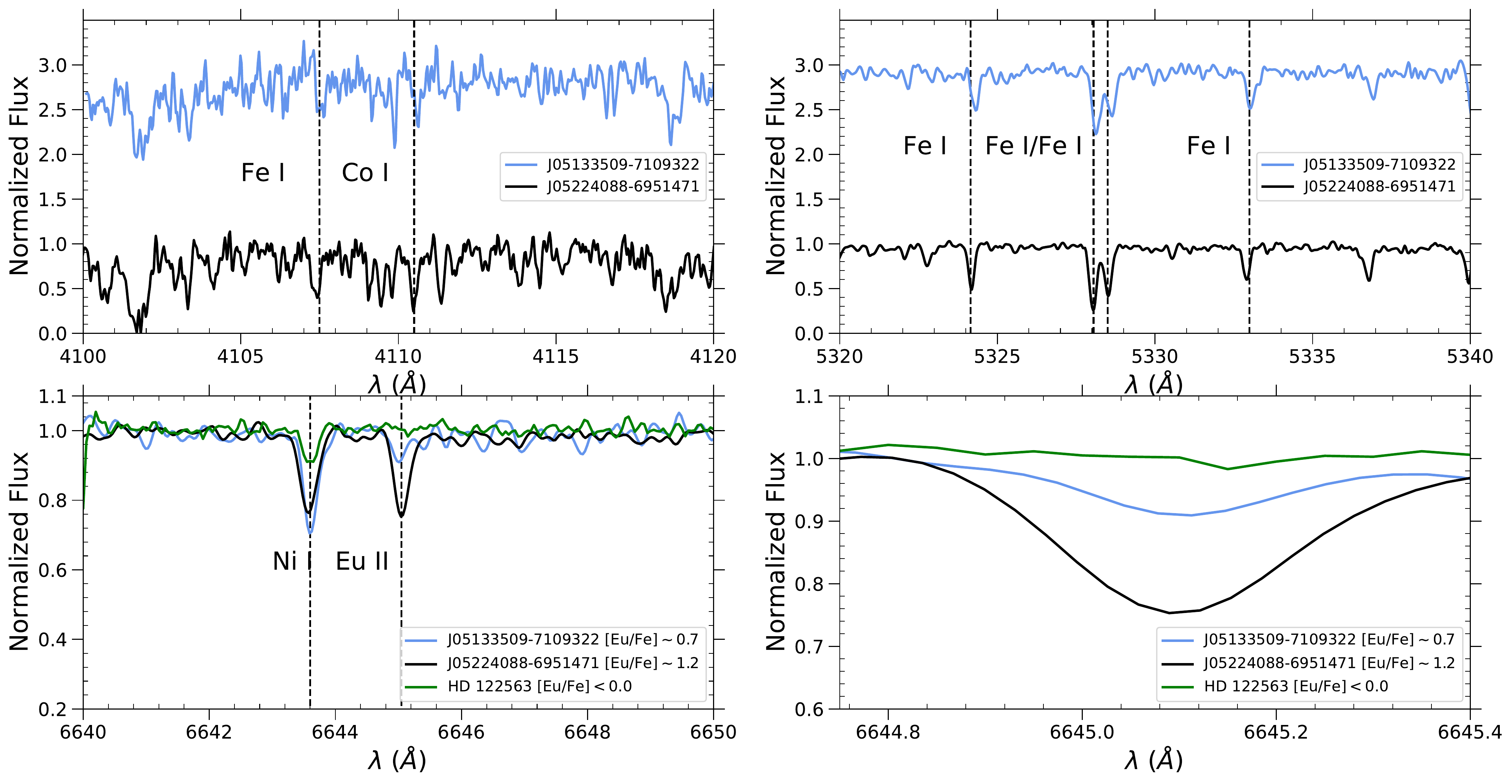}
\caption{Spectra for two LMC
giants in our sample 2MASS J05133509-7109322 ($[\text{Eu/Fe}] \approx
0.6$) and 2MASS J05224088-6951471 ($[\text{Eu/Fe}] \approx 1.5$) plus the
well-studied metal-poor giant HD 122563 ($[\text{Eu/Fe}] \lesssim 0$). 
All three stars have similar similar photospheric stellar parameters, and  
2MASS J05133509-7109322 \& 2MASS J05224088-6951471 
span our S/N range. Top left:
spectra in the blue in the wavelength range 4100 \AA~$\leq \lambda \leq$
4120 \AA.  Top right: spectra in the green in the wavelength range 5320
\AA~$\leq \lambda \leq$ 5340 \AA.  Bottom left: spectra in the red in the
wavelength range 6640 \AA~$\leq \lambda \leq$ 6650 \AA~in the vicinity
of the \ion{Eu}{2} line at 6645 \AA.  Bottom right: close-up view of
the \ion{Eu}{2} line at 6645 \AA.\label{europium_line}}
\end{figure*}

Our inferred abundances of the $r$-process element europium in
metal-poor Magellanic Cloud giants are remarkable.  Most extraordinary
are the $[\text{Eu/Fe}] \approx 1.5$ abundances observed in the
stars 2MASS J05121686-6517147 and 2MASS J05224088-6951471.  In Figure
\ref{europium_line} we show  europium lines at 6645 \AA~to illustrate
their large equivalent widths and therefore the large abundances of
europium in our sample.  The occurrence of so-called r-II stars is very
high in our sample of metal-poor Magellanic Cloud giants.  Indeed, three
of our nine metal-poor LMC giants are significantly enhanced in europium
according to the classical definition of r-II stars (i.e., $[\text{Eu/Fe}]
\ge 1.0$ and $[\text{Eu/Ba}] \geq 0$) from \citet{beers2005}.
If instead we adopt the new empirical r-II classification (i.e.,
$[\text{Eu/Fe}] \ge 0.7$) proposed by \citet{holmbeck2020}, then six
out of our nine metal-poor LMC giants are r-II stars (with two more at
$[\text{Eu/Fe}] = 0.65$ and and $[\text{Eu/Fe}] = 0.50$ very close to
the defined limit).  Regardless of which classification we adopt, all
nine of our LMC giants are considered r-I stars (i.e., $[\text{Eu/Fe}]
\ge 0.3$) and therefore $r$-process enhanced.  While we were only able
to infer europium abundances for two of our four metal-poor SMC giants,
one is an r-II star according to \citet{beers2005} while both are
r-II stars according to \citet{holmbeck2020}.  While we see elevated
$[\text{Eu/Fe}]$ abundances in the metallicity range $-2.6 \lesssim
[\text{Fe/H}] \lesssim -1.8$, at higher metallicities $[\text{Eu/Fe}]$
levels off close to the values observed in the Milky Way at similar
metallicities \citep[e.g.,][]{vanderswaelmen2013,kobayashi2020}.

Among the neutron-capture abundances we inferred, grids
of non-LTE corrections are only readily available for barium.
Non-LTE barium corrections for metal-poor giants are about $-0.1$ dex
\citep[e.g.,][]{amarsi2020}.

\section{Discussion}\label{discussion}

In both the Large and Small Magellanic Clouds, in the metallicity interval
$-2.6 \lesssim [\text{Fe/H}] \lesssim -2.0$ we observe high average
values of $[\text{Eu/Fe}]$, increased occurrences of $r$-process enhanced
stars relative to the Milky Way, and $\alpha$ element enhancements.
To quantify the significance of these first two observations, we compare
our results with the distribution of Milky Way halo $[\text{Eu/Fe}]$
abundances observed by \citet{barklem2005} and \citet{jacobson2015}
in the same range of metallicity.  Since our sample was selected
independently of $[\text{Eu/Fe}]$, it should provide an unbiased view of
the Magellanic Clouds' $[\text{Eu/Fe}]$ distribution in the metallicity
range $-2.6 \lesssim [\text{Fe/H}] \lesssim -1.5$.  Likewise, the stars
observed by \citet{barklem2005} and \citet{jacobson2015} were selected
without regard to their $[\text{Eu/Fe}]$ abundances, so the union of
the \citet{barklem2005} and \citet{jacobson2015} catalogs provides
an unbiased view of the distribution of $[\text{Eu/Fe}]$ in Milky
Way halo stars in the metallicity range $-2.6 \lesssim [\text{Fe/H}]
\lesssim -1.5$.  Because all of the stars in our Magellanic Clouds
sample have $[\text{Eu/Ba}] \geq 0$ as expected from $r$-process
nucleosynthesis, we first remove all stars from the \citet{barklem2005}
and \citet{jacobson2015} catalogs with $[\text{Eu/Ba}] < 0$.\footnote{This
requirement forces us to exclude stars with only upper limits on barium or
europium abundances, but the inclusion of stars with upper limits would
make the inferences we describe in the next paragraph even stronger.}
From here, we refer to this sample as our control sample.

For each of the nine stars in our LMC sample, we calculate the probability
$p_{i}$ that a randomly selected star from the control sample with
$|[\text{Fe/H}]_{\text{LMC}} - [\text{Fe/H}]_{\text{control}}|
< \sigma_{[\text{Fe/H}]}$ has $[\text{Eu/Fe}]_{\text{control}}
\ge [\text{Eu/Fe}]_{\text{LMC}}$.  We then use these individual
probabilities to calculate the probability $9! \prod_{i=1}^{9} p_{i}$
that a nine-star sample from the control sample has a more extreme
$[\text{Eu/Fe}]$ distribution than what we observe in the LMC.  We find
that random sampling from the control $[\text{Eu/Fe}]$ distribution
has only a 1 in 20600 chance of producing the
extreme $[\text{Eu/Fe}]$ we observe in the LMC, equivalent to a
3.90-$\sigma$ offset.  A similar calculation
for the two SMC stars for which were able to infer $[\text{Eu/Fe}]$
indicates that random sampling from the control $[\text{Eu/Fe}]$
distribution has only a 1 in 279 chance of producing the
extreme $[\text{Eu/Fe}]$ distribution we observe in the SMC, equivalent
to a 2.69-$\sigma$ offset.  Overall, we find
that random sampling from the control $[\text{Eu/Fe}]$ distribution
has only a 1 in $2.88 \times 10^{6}$ 
chance of producing the extreme $[\text{Eu/Fe}]$ distribution we
observe in the metal-poor Magellanic Clouds sample, equivalent to a
4.96-$\sigma$ offset.

All 11 metal-poor Magellanic Cloud giants for which we could infer a
europium abundance would be classified as r-I or r-II stars according
to the criteria proposed by \citet{beers2005}.  We find that three of
nine metal-poor LMC giants would be classified as r-II stars according
to \citet{beers2005}; six would be classified as r-II stars according
to \citet{holmbeck2020}.  Both metal-poor SMC giants for which
we could infer europium abundances would be considered r-II stars
according to \citet{holmbeck2020}; only one would be considered an
r-II star according to \citet{beers2005}.  If we model the occurrence
of $r$-process enhanced stars with a binomial distribution and use an
uninformative Beta(1,1) prior in a Bayesian analysis, then the posterior
on occurrence will have a Beta distribution as well.  We thereby find
in the metallicity range $-2.6 \lesssim [\text{Fe/H}] \lesssim -1.5$ r-I
occurrence rates in the LMC, SMC, and complete Magellanic Cloud samples
to be $93^{+5}_{-10}$\%, $79^{+15}_{-25}$\%, and $94^{+4}_{-9}$\%.
According to the \citet{beers2005} definition, we find in the same
metallicity range r-II occurrence rates in the LMC, SMC, and complete
Magellanic Cloud samples to be $36^{+15}_{-14}$\%, $50^{+25}_{-25}$\%,
and $38^{+14}_{-13}$\%.  According to the \citet{holmbeck2020} definition,
we find in the same metallicity range r-II occurrences in the LMC,
SMC, and complete Magellanic Cloud samples to be $64^{+14}_{-15}$\%,
$79^{+15}_{-25}$\%, and $70^{+12}_{-14}$\%.

The occurrence rates of $r$-process enhanced stars we observe in
the Magellanic Clouds are significantly higher than in the halo of
the Milky Way or most ultra-faint dSph galaxies.  \citet{barklem2005}
found that 14\%/3\% of metal-poor Milky Way halo stars are classified as
r-I/r-II according to the \citet{beers2005} definitions.  More recent
estimates by the $r$-process Alliance have found similar occurrences
\citep[e.g.,][]{hansen2018,sakari2018}.  We note that while
\citet{ezzeddine2020} report somewhat higher rates of $r$-process
enhancement, even their possibly biased input sample still produces
significantly lower occurrences of both r-I and r-II stars than what
we observe in the Magellanic Clouds.  The ultra-faint dSph galaxies
Reticulum II and Tucana III are the only environments in which
similar occurrences of $r$-process enhanced stars have been found
\citep[e.g.,][]{ji2016a,ji2016b,roederer2016,hansen2017,marshall2019}.

In addition to their enhancements in $[\text{Eu/Fe}]$, in the
metallicity range $-2.6 \lesssim [\text{Fe/H}] \lesssim -2.0$ all of
our metal-poor giants in the LMC and SMC giants are also enhanced in
the $\alpha$ elements.  These elevated $[\alpha/\text{Fe}]$ abundances
are indicative of formation by a time at which thermonuclear supernovae
had not yet started contributing significantly to the Magellanic Clouds'
stellar populations' chemical evolution \citep[e.g.,][]{tinsley1979}.
The implication is that the metal-poor LMC and SMC giants in our
sample formed after the Magellanic Clouds' first core-collapse
supernovae but before thermonuclear supernovae could contribute
significantly to the Magellanic Clouds' chemical evolution.  The
copious iron-peak nucleosynthesis produced in thermonuclear supernovae
causes $[\alpha/\text{Fe}]$ to decline, and the metallicity at which
$[\alpha/\text{Fe}]$ transitions from a high and constant value to a
linearly decreasing function of metallicity is often called the ``knee''
in the $[\text{Fe/H}]$---$[\alpha/\text{Fe}]$ plane.

While the scenario outlined above is likely too simplistic and the
metallicity at which $[\alpha/\text{Fe}]$ declines depends on stellar
and supernovae feedback, outflows, and inflows, $[\alpha/\text{Fe}]$
tends to begin its decline at higher metallicities in more massive
galaxies \citep[e.g.,][]{suda2017,nidever2020}.  The Magellanic Clouds
appear to be an outlier in this regard, as it has been argued that
in the Magellanic Clouds $[\alpha/\text{Fe}]$ begins its decline below
$[\text{Fe/H}] \approx -2.2$ \citep{nidever2020}.  These upper limits are
below the metallicities at which $[\alpha/\text{Fe}]$ begins to decline
in less massive galaxies like Sagittarius, Fornax, and Sculptor, though
we acknowledge that the $\alpha$-abundance plateaus in these galaxies are
are not as clearly defined as in the Milky Way \citep[e.g.][]{kirby2011b}.
Indeed, if the Magellanic Clouds followed the trends with mass observed
in Sagittarius, Fornax, and Sculptor then $[\alpha/\text{Fe}]$ should
start to decline in the Magellanic Clouds at $[\text{Fe/H}] \approx -1.2$
\citep[e.g.,][]{nidever2020}.

\begin{figure*}
\plotone{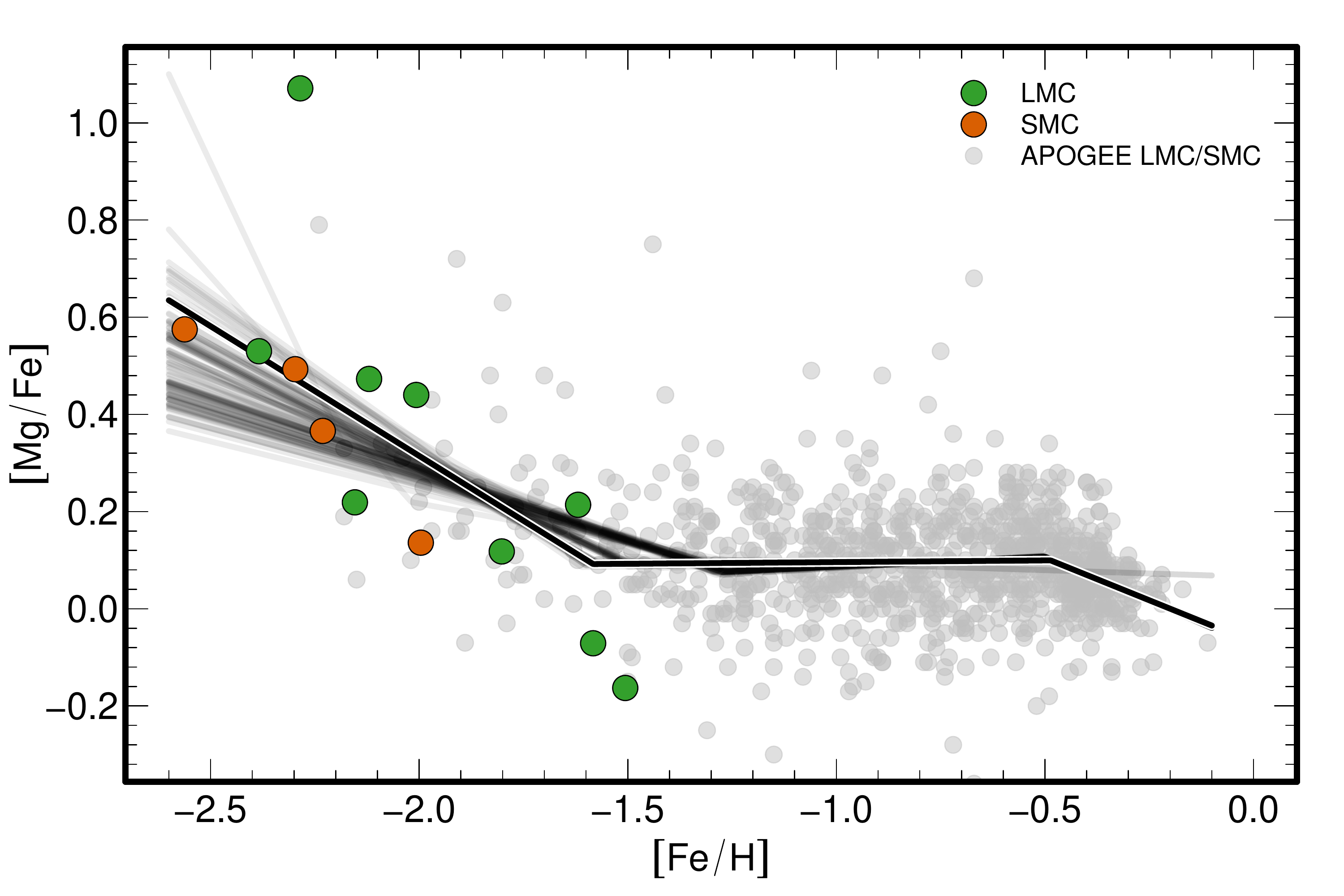}
\caption{Evolution of $[\text{Mg/Fe}]$ with $[\text{Fe/H}]$ in the
Magellanic Clouds.  We plot as dark green circles our nine LMC giants
and as blue circles our four SMC giants.  We plot as gray circles LMC
and SMC giants observed by SDSS-IV/APOGEE-2 that are part of SDSS DR16
\citep[e.g.,][]{nidever2020}.  The opaque black line is the best-fit
three-segment linear model for our point estimates of $[\text{Fe/H}]$ and
$[\text{Mg/Fe}]$.  The transparent lines are the best-fit three-segment
linear models for each Monte Carlo iteration accounting for the
uncertainties in our $[\text{Fe/H}]$ and $[\text{Mg/Fe}]$ inferences.
For our point estimates, the fitting procedure identified changes in
slope at $[\text{Fe/H}] \approx -1.6$ and $[\text{Fe/H}] \approx -0.49$.
The results of the Monte Carlo simulation are consistent with
this inference. The locations and slopes of the best-fit segments make
it difficult to interpret the relationship between $[\text{Fe/H}]$
and $[\text{Mg/Fe}]$ in the Magellanic Clouds in the classical
\citet{tinsley1979} picture.  We are therefore unable to claim a detection
of the knee in Magellanic Clouds' $[\text{Fe/H}]$---$[\text{Mg/Fe}]$
plane based on these data.\label{knee_fig}}
\end{figure*}

Our stars extend to lower metallicities than most of the Magellanic
Cloud stars studied by \citet{nidever2020}, and we therefore attempted
to locate the metallicity at which $[\alpha/\text{Fe}]$ begins to
decline in the Magellanic Clouds.  To algorithmically search for the
knee in the Magellanic Clouds' $[\text{Fe/H}]$---$[\alpha/\text{Fe}]$
distribution, we use the \texttt{segmented} \citep{muggeo2003}
package in \texttt{R} \citep{r20} to fit a segmented piece-wise linear
model to the $[\text{Fe/H}]$ and $[\text{Mg/Fe}]$ values presented
in Figure \ref{alphas_fig}. For our point estimates, the best fit piece-wise
linear model features a declining $[\text{Mg/Fe}]$ trend in the
$[\text{Fe/H}]$---$[\text{Mg/Fe}]$ plane, even at $[\text{Fe/H}] \approx
-2.6$.  We were unable to identify a plateau at constant $[\text{Mg/Fe}]$
at low metallicity using the combination of our observations and SDSS
DR16 data, so it is difficult to interpret this result according
to the \citet{tinsley1979} paradigm.  We also used a Monte
Carlo simulation to investigate the impact of our $[\text{Fe/H}]$
and $[\text{Mg/Fe}]$ uncertainties on our inferences about the knee.
On each iteration of our Monte Carlo simulation, we sampled values
of $[\text{Fe/H}]$ and $[\text{Mg/Fe}]$ from normal distributions
assuming the uncertainties given in Table \ref{chem_abundances}.
We then fit a segmented piece-wise linear model to the $[\text{Fe/H}]$
and $[\text{Mg/Fe}]$ values produced in each iteration and saved the
result.  While our $[\text{Fe/H}]$ and $[\text{Mg/Fe}]$ uncertainties
limit our ability to precisely infer segment slopes and end points,
we never observed a plateau at constant $[\text{Mg/Fe}]$ at low
metallicity. We plot the results of these calculations in
Figure \ref{knee_fig}.

We now propose a scenario that explains the ubiquitous $r$-process
enhancement of the Magellanic Clouds' $\alpha$-enhanced metal-poor
stars and accommodates what is known about the Magellanic Clouds' other
properties.  In accord with our estimates of their ages, the $\alpha$
and $r$-process enhanced metal-poor giants we observe in both the LMC and
SMC formed in each galaxy's initial burst of metal-poor star formation
more than 10 Gyr in the past \citep[e.g.,][]{harris2004,harris2009}.
Relative to the onset of this initial burst of star formation, they must
have formed after a few Myr but before a few tens to 100 Myr to explain
our observation of their $[\alpha/\text{Fe}] \approx 0.4$.  While the
Milky Way was able to produce stars with $[\text{Fe/H}] \approx -1.0$ and
$[\alpha/\text{Fe}] \approx 0.4$, the chemical evolution of the Magellanic
Clouds moved more slowly such that even stars with $[\text{Fe/H}] \lesssim
-2$ show a decline in $[\alpha/\text{Fe}]$ with increasing $[\text{Fe/H}]$
\citep[e.g., Figure \ref{knee_fig} and][]{nidever2020}.

We propose two reasons for this apparently slow chemical evolution.
The first was the Magellanic Clouds' ongoing accretion of unenriched
gas from the cosmic web, extending the duration of their metal-poor
star formation as the supply of unenriched gas was sufficient to keep
$[\text{Fe/H}] \lesssim -2$.  Unlike most of the Milky Way's satellite
galaxy population, the Magellanic Clouds' long history of evolution in
isolation protected them from ram-pressure stripping and strangulation,
consequently ensuring a consistent supply of unenriched gas from the
cosmic web.  The second reason for the Magellanic Clouds' extended era of
metal-poor star formation was that they could not be quenched by either
stellar and supernova feedback (because of their relatively high masses
compared to the classical and ultra-faint dwarf galaxies) or AGN feedback
(because of their low masses compared to the Milky Way and M31).

The Magellanic Clouds' extended durations of metal-poor star formation
combined with their high star formation rates (relative to the Milky
Way's classical and ultra-faint dSph galaxies) ensured that even
nucleosynthesis that occurs in uncommon astrophysical events would
contribute to their chemical evolution.  In particular, the ubiquitous
$r$-process enhancement of the Magellanic Clouds' $\alpha$-enhanced stars
we observe was the result of (1) rare nucleosynthetic events that (2) have
characteristic timescale longer than the core-collapse supernova timescale
but shorter than or comparable to the thermonuclear supernova timescale.
The former fact is necessary to accommodate the rarity of $r$-process
enhanced ultra-faint dSph galaxies, while the latter fact is necessary to
accommodate the relative rarity of $r$-process enhanced stars in the halo
of the Milky Way.  Because rare classes of core-collapse supernovae like
collapsars or magnetorotationally driven supernovae would occur quickly,
our observations favor compact object mergers involving a neutron star as
the origin of the $r$-process elements observed in $r$-process enhanced
stars at low metallicities in the Milky Way, the Magellanic Clouds,
and Reticulum II.

\begin{figure*}
\plotone{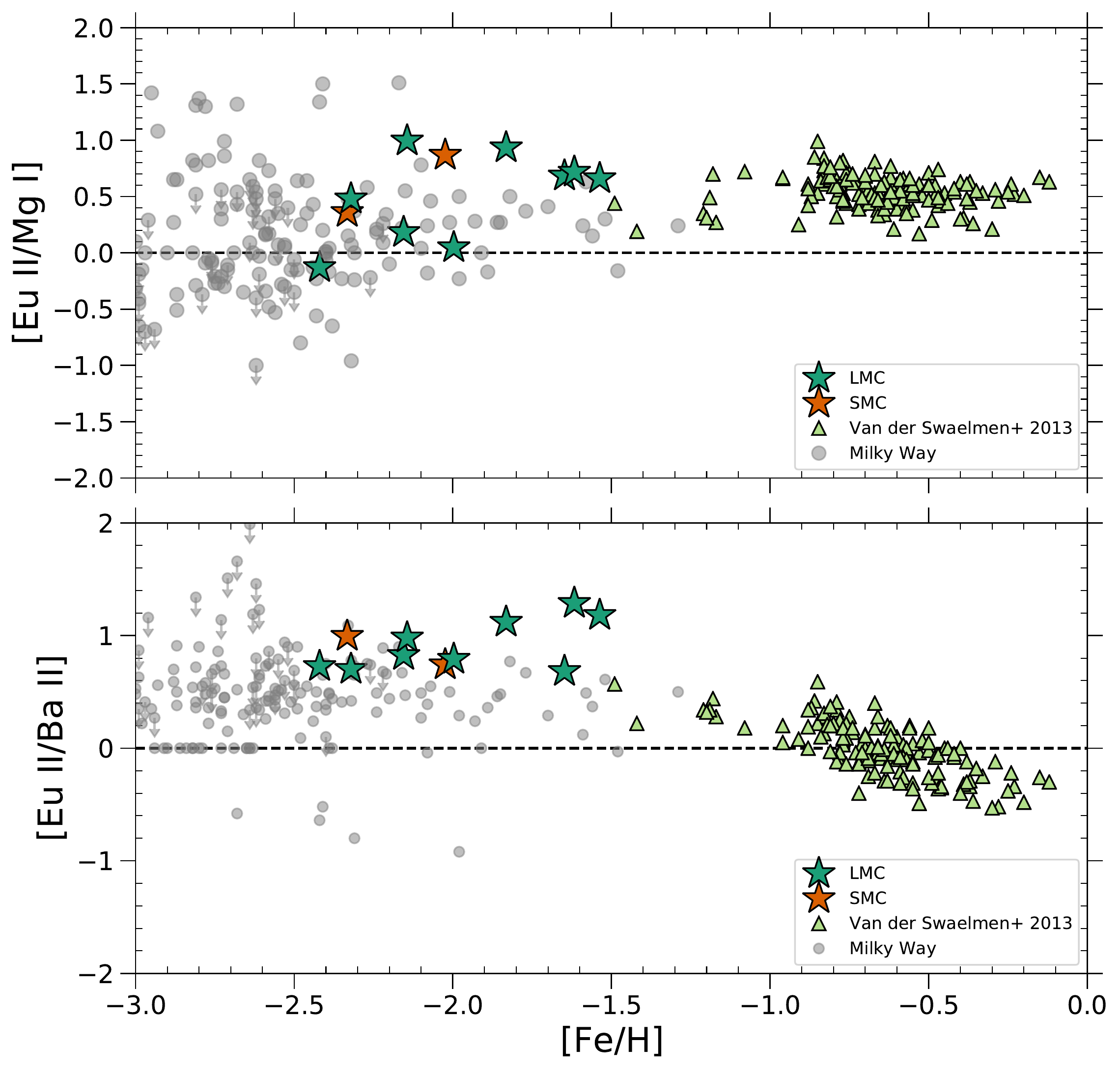}
\caption{Magellanic Cloud $[\text{Eu/Mg}]$ and $[\text{Eu/Ba}]$
abundances as a function of metallicity.  We plot as dark green stars
our nine LMC giants and as orange stars our two SMC giants for which
we were able to infer europium abundances.  We plot as light green
triangles LMC bar giants from \citet{vanderswaelmen2013}.  We plot
as gray circles metal-poor Milky Way giants from \citet{barklem2005}
and \citet{jacobson2015} selected for high-resolution follow-up
observations without regard to their $r$-process abundances.  We find that
$[\text{Eu/Mg}]$ is has no dependence on metallicity in the range $-2.6
\lesssim [\text{Fe/H}] \lesssim 0$ in the Magellanic Clouds.  On the
other hand, $[\text{Eu/Ba}]$ declines with metallicity.  The first
trend is consistent with a constant ratio of the nucleosynthesis of
magnesium in core-collapse supernovae and europium in compact object
mergers involving a neutron star at times in excess of 100 Myr after
the onset of star formation.  The second trend is consistent with a
constant contribution of $r$-process barium from compact object mergers
involving a neutron star combined with increasing contributions with
time (and therefore metallicity) of $s$-process barium from low-mass
AGB stars.\label{eumgxfeh}}
\end{figure*}

We argue that our observation of the Magellanic Clouds'
$[\alpha/\text{Fe}]$ and $[\text{Eu/Fe}]$ distributions in the
metallicity interval $-2.6 \lesssim [\text{Fe/H}] \lesssim -2.0$ favors
the nucleosynthesis of $r$-process elements with a delay between a
few Myr and a few tens to 100 Myr after the onset of star formation.
In accord with our results for the Magellanic Clouds, \citet{matsuno2021}
used simple chemical evolution models to suggest that the progenitor
Gaia-Enceladus supports the idea of delayed $r$-process enrichment with a
timescale shorter than the characteristic thermonuclear supernovae delay
time.  Likewise, our upper limit on the delay time is in agreement with
the conclusions of \citet{skuladottir2020}.  Based on the relationship
between the $r$-process abundances and the star formation histories of
classical dSph galaxies and the Milky Way's solar neighborhood, those
authors proposed the existence of two different $r$-process sources:
a quick source that occurs less than 100 Myr after the onset of star
formation and a delayed source that occurs more than 4 Gyr after the onset
of star formation.  In contrast with their results, as we plot in Figure
\ref{eumgxfeh} the $[\text{Eu/Mg}]$ ratios of the Magellanic Clouds
are constant at supersolar values throughout the entire metallicity
range we examined.  We argue that this trend is consistent with a
constant ratio of the nucleosynthesis of magnesium in core-collapse
supernovae and europium in compact object mergers involving a neutron
star at times in excess of 100 Myr after the onset of star formation.
Although the Magellanic Clouds' $[\text{Eu/Mg}]$ abundances are flat,
we show in Figure \ref{eumgxfeh} that there is a clear declining trend
in $[\text{Eu/Ba}]$ abundances with increasing metallicity that we
attribute to the increasing importance with metallicity of $s$-process
barium produced in low-mass asymptotic giant branch (AGB) stars.

We predict that future observations of galaxies that resemble the
Magellanic Clouds in two key properties will also display metal-poor
stellar populations with ubiquitous $r$-process enhancement.  The first
key property is evolution substantially in isolation that allows a galaxy
to accrete unenriched gas from the cosmic web and avoid ram-pressure
stripping/strangulation inside the virial radius of a more massive halo.
The second key property is an intermediate mass that is high enough to
be robust against stellar and supernovae feedback but low enough not to
be significantly affected by AGN feedback or accretion shocks.

\section{Conclusion}\label{conclusion}

We observed with high-resolution Magellan/MIKE spectroscopy nine LMC and
four SMC stars selected using the mid-infrared metal-poor star selection
of \citet{schlaufman2014} and archival data.  These stars are the most
metal-poor Magellanic Cloud stars yet subject to a comprehensive abundance
analysis.  We find that in the interval $-2.6 \lesssim [\text{Fe/H}]
\lesssim -1.5$ these stars are similar to Milky Way halo stars in their
$\alpha$, light odd-$Z$, iron-peak, and $s$-process neutron-capture
element abundances, including their enhancement in $\alpha$ elements
magnesium, calcium, and titanium.  We discover that both the Large and
Small Magellanic Clouds are ubiquitously enhanced in the $r$-process
element europium relative to the Milky Way's halo.  The probabilities that
the large $[\text{Eu/Fe}]$ abundances we observe in the LMC, the SMC,
and our complete Magellanic Clouds sample could be explained by random
sampling from an $[\text{Eu/Fe}]$ distribution like the Milky Way's are
less than 1 in 20600, 1 in 279, and 1 in $2.88 \times 10^{6}$.  These probabilities
would be equivalent to 3.90, 2.69, and
4.96 $\sigma$ in a Gaussian distribution.  Even though we studied the
Magellanic Clouds' $[\alpha/\text{Fe}]$---$[\text{Fe/H}]$ distribution
to unprecedentedly low metallicities, we could not to identify a
plateau in the $[\alpha/\text{Fe}]$---$[\text{Fe/H}]$  distribution and
therefore were unable to identify the $[\text{Fe/H}]$ value at which
$[\alpha/\text{Fe}]$ begins to decrease (i.e., the ``knee'').

We argue the ubiquitous $\alpha$ and europium enhancements observed in
the Magellanic Clouds are a product of their isolated chemical evolution
and long history of accretion from the cosmic web that extended the era of
metal-poor star formation for a much longer time in the Magellanic Clouds
than in the Milky Way or M31.  This extended duration of star formation
allowed time for $r$-process nucleosynthesis in events that began to
occur somewhere between the core-collapse supernova timescale (a few
Myr after the onset of star formation) and the thermonuclear supernova
timescale (a few tens to 100 Myr after the onset of star formation).
These events that produced the europium enhancements we observe in the
Magellanic Clouds must have been rare, otherwise $r$-process enhanced
stars would be ubiquitous in the Milky Way's ultra-faint dSphs. 
Compact object mergers involving a neutron star are the best candidate
for a rare event that produces $r$-process nucleosynthesis on a timescale
longer than the core-collapse supernova timescale.  Our observations
provide strong support for compact object mergers involving a neutron
star occurring after core-collapse supernova but before thermonuclear
supernova as the site of the $r$-process nucleosynthesis responsible for
the ubiquitous europium enhancement we observe in the Magellanic Clouds in
the interval $-2.6 \lesssim [\text{Fe/H}] \lesssim -1.5$.  We predict that
metal-poor stars in intermediate-mass galaxies that evolved substantially
in isolation will also be enhanced in $r$-process elements.

\clearpage
\begin{acknowledgments}

We thank the referee for the valuable comments that helped us improve
our paper.  We thank David Nataf for his help with a discussion on the
extinction towards the Magellanic system.  Andrew R.\ Casey is supported
in part by the Australian Research Council through a Discovery Early
Career Researcher Award (DE190100656).  Parts of this research were
supported by the Australian Research Council Centre of Excellence for
All Sky Astrophysics in 3 Dimensions (ASTRO 3D), through project number
CE170100013.  Joshua D.\ Simon is supported by NSF grant AST-1714873.
Alexander P.\ Ji acknowledges support from a Carnegie Fellowship
and the Thacher Research Award in Astronomy.  This work is based in
part on observations made with the Spitzer Space Telescope, which is
operated by the Jet Propulsion Laboratory, California Institute of
Technology under a contract with NASA.  This publication makes use
of data products from the Wide-field Infrared Survey Explorer, which
is a joint project of the University of California, Los Angeles, and
the Jet Propulsion Laboratory/California Institute of Technology, and
NEOWISE, which is a project of the Jet Propulsion Laboratory/California
Institute of Technology.  WISE and NEOWISE are funded by the National
Aeronautics and Space Administration.  This publication makes use of
data products from the Two Micron All Sky Survey, which is a joint
project of the University of Massachusetts and the Infrared Processing
and Analysis Center/California Institute of Technology, funded by the
National Aeronautics and Space Administration and the National Science
Foundation.  This research has made use of the NASA/IPAC Infrared
Science Archive, which is funded by the National Aeronautics and Space
Administration and operated by the California Institute of Technology.
This paper includes data gathered with the 6.5 m Magellan Telescopes
located at Las Campanas Observatory, Chile.  Australian access to the
Magellan Telescopes was supported through the National Collaborative
Research Infrastructure Strategy of the Australian Federal Government.
The national facility capability for SkyMapper has been funded through ARC
LIEF grant LE130100104 from the Australian Research Council, awarded to
the University of Sydney, the Australian National University, Swinburne
University of Technology, the University of Queensland, the University
of Western Australia, the University of Melbourne, Curtin University of
Technology, Monash University and the Australian Astronomical Observatory.
SkyMapper is owned and operated by The Australian National University's
Research School of Astronomy and Astrophysics.  The survey data were
processed and provided by the SkyMapper Team at ANU.  The SkyMapper
node of the All-Sky Virtual Observatory (ASVO) is hosted at the National
Computational Infrastructure (NCI).  Development and support the SkyMapper
node of the ASVO has been funded in part by Astronomy Australia Limited
(AAL) and the Australian Government through the Commonwealth's Education
Investment Fund (EIF) and National Collaborative Research Infrastructure
Strategy (NCRIS), particularly the National eResearch Collaboration Tools
and Resources (NeCTAR) and the Australian National Data Service Projects
(ANDS).  Funding for the Sloan Digital Sky Survey IV has been provided by
the Alfred P.\ Sloan Foundation, the U.S.  Department of Energy Office of
Science, and the Participating Institutions.  SDSS-IV acknowledges support
and resources from the Center for High Performance Computing  at the
University of Utah.  The SDSS website is \url{www.sdss.org}.  SDSS-IV is
managed by the Astrophysical Research Consortium for the Participating
Institutions of the SDSS Collaboration including the Brazilian
Participation Group, the Carnegie Institution for Science, Carnegie
Mellon University, Center for Astrophysics | Harvard \& Smithsonian, the
Chilean Participation Group, the French Participation Group, Instituto de
Astrof\'isica de Canarias, Johns Hopkins University, Kavli Institute for
the Physics and Mathematics of the Universe (IPMU) / University of Tokyo,
the Korean Participation Group, Lawrence Berkeley National Laboratory,
Leibniz Institut f\"ur Astrophysik Potsdam (AIP),  Max-Planck-Institut
f\"ur Astronomie (MPIA Heidelberg), Max-Planck-Institut f\"ur Astrophysik
(MPA Garching), Max-Planck-Institut f\"ur Extraterrestrische Physik
(MPE), National Astronomical Observatories of China, New Mexico State
University, New York University, University of Notre Dame, Observat\'ario
Nacional / MCTI, The Ohio State University, Pennsylvania State University,
Shanghai Astronomical Observatory, United Kingdom Participation Group,
Universidad Nacional Aut\'onoma de M\'exico, University of Arizona,
University of Colorado Boulder, University of Oxford, University of
Portsmouth, University of Utah, University of Virginia, University of
Washington, University of Wisconsin, Vanderbilt University, and Yale
University.  This work has made use of data from the European Space
Agency (ESA) mission {\it Gaia} (\url{https://www.cosmos.esa.int/gaia}),
processed by the {\it Gaia} Data Processing and Analysis Consortium
(DPAC, \url{https://www.cosmos.esa.int/web/gaia/dpac/consortium}).
Funding for the DPAC has been provided by national institutions, in
particular the institutions participating in the {\it Gaia} Multilateral
Agreement.  This research has made use of ``Aladin sky atlas'' developed
at CDS, Strasbourg Observatory, France \citep{bonnarel2000,boch2014}.
The Digitized Sky Survey was produced at the Space Telescope Science
Institute under U.S. Government grant NAG W-2166.  The images of these
surveys are based on photographic data obtained using the Oschin Schmidt
Telescope on Palomar Mountain and the UK Schmidt Telescope.  The plates
were processed into the present compressed digital form with the
permission of these institutions.  The National Geographic Society-Palomar
Observatory Sky Atlas (POSS-I) was made by the California Institute of
Technology with grants from the National Geographic Society.  The Second
Palomar Observatory Sky Survey (POSS-II) was made by the California
Institute of Technology with funds from the National Science Foundation,
the National Aeronautics and Space Administration, the National Geographic
Society, the Sloan Foundation, the Samuel Oschin Foundation, and the
Eastman Kodak Corporation.  The Oschin Schmidt Telescope is operated by
the California Institute of Technology and Palomar Observatory.  The UK
Schmidt Telescope was operated by the Royal Observatory Edinburgh, with
funding from the UK Science and Engineering Research Council (later the
UK Particle Physics and Astronomy Research Council), until 1988 June,
and thereafter by the Anglo-Australian Observatory.  The blue plates of
the southern Sky Atlas and its Equatorial Extension (together known as the
SERC-J), the near-IR plates (SERC-I), as well as the Equatorial Red (ER),
and the Second Epoch [red] Survey (SES) were all taken with the UK Schmidt
telescope at the AAO.  This research has made use of NASA's Astrophysics
Data System.  This research has made use of the SIMBAD database, operated
at CDS, Strasbourg, France \citep{wenger2000}.  This research has made
use of the VizieR catalogue access tool, CDS, Strasbourg, France (DOI:
10.26093/cds/vizier).  The original description of the VizieR service
was published in 2000, A\&AS 143, 23 \citep{ochsenbein2000}.

\end{acknowledgments}

\vspace{5mm}
\facilities{CTIO:2MASS, Magellan:Clay (MIKE echelle spectrograph), IRSA,
Skymapper, Spitzer (IRAC), WISE.}

\software{\texttt{astropy} \citep{astropy2013,astropy2018},
          \texttt{CarPy} \citep{kelson2000,kelson2003},
          \texttt{q2} \citep{ramirez2014},
          \texttt{isochrones} \citep{morton2015},
          \texttt{numpy} \citep{vanderwalt2011},
          \texttt{MultiNest} \citep{feroz2008,feroz2009,feroz2019},
          \texttt{pandas} \citep{scipy2010},
          \texttt{R} \citep{r20},
          \texttt{scipy} \citep{scipy2020}
          \texttt{IRAF} \citep{iraf1986,iraf1993}
          }

\clearpage
\bibliography{ms09}{}
\bibliographystyle{aasjournal}

\begin{longrotatetable} 
\begin{deluxetable*}{lcccccc} 
\tablecaption{Stellar Properties and Adopted Parameters\label{stellar_params}} 
\tablewidth{0pt} 
\tablehead{ 
\colhead{} & \colhead{} & \colhead{} & \colhead{} & \colhead{} & \colhead{} & \colhead{}} 
\startdata
Property & J00251849-7140074 & J00262394-7128549 & J00263959-7122102 & J00273753-7125114 & & Units \\
         &               SMC &               SMC &               SMC &               SMC & & \\
\hline
\textbf{Photometric Properties} & & & & & \\
SkyMapper $g$ & $16.575\pm0.036$ & $17.604\pm0.016$ & $16.801\pm0.025$ & $16.987\pm0.028$ & & AB mag \\
SkyMapper $r$ & $15.917\pm0.020$ & $16.980\pm0.014$ & $16.172\pm0.013$ & $16.107\pm0.013$ & & AB mag \\
SkyMapper $i$ & $15.356\pm0.003$ & $16.403\pm0.017$ & $15.576\pm0.040$ & $15.475\pm0.004$ & & AB mag \\
SkyMapper $z$ & $15.171\pm0.009$ & $16.232\pm0.017$ & $15.407\pm0.009$ & $15.208\pm0.003$ & & AB mag \\
2MASS $J$ & $13.938\pm0.026$ & $15.076\pm0.045$ & $14.236\pm0.030$ & $14.038\pm0.029$ & & Vega mag \\
2MASS $H$ & $13.327\pm0.035$ & $14.519\pm0.060$ & $13.580\pm0.026$ & $13.351\pm0.030$ & & Vega mag \\
2MASS $K_{\text{s}}$ & $13.119\pm0.037$ & $14.295\pm0.078$ & $13.462\pm0.051$ & $13.242\pm0.040$ & & Vega mag \\
WISE $W1$ & $13.106\pm0.024$ & $14.247\pm0.026$ & $13.423\pm0.024$ & $13.092\pm0.024$ & & Vega mag \\
WISE $W2$ & $13.104\pm0.026$ & $14.268\pm0.038$ & $13.451\pm0.028$ & $13.064\pm0.025$ & & Vega mag \\
\textbf{Stellar Properties} & & & & & \\
Luminosity\tablenotemark{a} $L_{\ast}$ & $1507^{+12}_{-22}$ & $954^{+10}_{-15}$ &  $1387\pm10$ & $1577\pm10$ & & L$_{\odot}$ \\
Radius\tablenotemark{a} $R_{\ast}$ & $68\pm1$ & $53\pm1$ &  $61\pm1$ & $71\pm1$ & & R$_{\odot}$ \\
Distance\tablenotemark{a} $d_{\text{iso}}$ & $55.0^{+0.2}_{-0.4}$ & $72.1^{+0.3}_{-0.5}$    & $59.6^{+0.2}_{-0.2}$ & $58.3^{+0.1}_{-0.1}$ & & kpc \\
Mass\tablenotemark{a} $M_{\odot}$ & $0.77\pm0.01$ & $0.76\pm0.01$ & $0.78\pm0.01$ & $0.76\pm0.01$ & & M$_{\odot}$ \\
Age\tablenotemark{a} $\tau$ & $13.2^{+0.2}_{-0.4}$ & $13.0^{+0.3}_{-0.6}$    & $13.4\pm0.1$ & $13.4\pm0.1$ & & Gyr \\
Extinction\tablenotemark{a} $A_{V}$ & $0.086^{+0.002}_{-0.003}$ & $0.086^{+0.001}_{-0.002}$  & $0.094^{+0.001}_{-0.002}$ & $0.099\pm0.001$ & & mag \\
Effective temperature\tablenotemark{a} $T_{\text{eff}}$ & $4372^{+5}_{-3}$ & $4388^{+8}_{-6}$ & $4481^{+2}_{-1}$ & $4326\pm10$ & & K \\
Surface gravity\tablenotemark{a} $\log{g}$ & $0.65\pm0.01$ & $0.87\pm0.01$ & $0.73\pm0.01$ & $0.61\pm0.01$ & & cm s$^{-2}$ \\
Metallicity $[\text{Fe/H}]$ & $-2.33\pm0.30$ & $-2.02\pm0.32$ & $-2.60\pm0.39$ & $-2.27\pm0.26$ &  & \\
Microturbulence $\xi$ & $2.71\pm0.50$ & $2.51\pm0.80$ & $3.01\pm0.45$ & $3.83\pm0.95$ & & km s$^{-1}$ \\
\hline
\hline
Property & J05121686-6517147 & J05133509-7109322 & J05141665-6454310 & J05143154-6505189 & J05150380-6647003 & Units  \\
         &               LMC &               LMC &               LMC &               LMC &               LMC &  \\
\hline
\textbf{Photometric Properties} & & & & &  & \\
SkyMapper $g$ & $17.329\pm0.266$ & $17.367\pm0.043$ & $17.556\pm0.021$ & $17.533\pm0.017$ & $17.138\pm0.014$ & AB mag  \\
SkyMapper $r$ & $16.019\pm0.099$ & $16.682\pm0.005$ & $17.056\pm0.007$ & $16.789\pm0.009$ & $16.247\pm0.009$ & AB mag  \\
SkyMapper $i$ & $15.265\pm0.063$ & $16.178\pm0.090$ & $16.721\pm0.043$ & $16.262\pm0.012$ & $15.635\pm0.007$ & AB mag  \\
SkyMapper $z$ & $15.017\pm0.038$ & $16.009\pm0.016$ & $16.577\pm0.017$ & $16.064\pm0.005$ & $15.417\pm0.012$ & AB mag  \\
2MASS $J$ & $13.584\pm0.030$ & $14.704\pm0.037$ & $15.363\pm0.064$ & $14.847\pm0.040$ & $14.063\pm0.030$ & Vega mag  \\
2MASS $H$ & $12.751\pm0.027$ & $13.976\pm0.051$ & $14.871\pm0.083$ & $14.070\pm0.040$ & $13.233\pm0.025$ & Vega mag  \\
2MASS $K_{\text{s}}$ & $12.584\pm0.032$ & $13.896\pm0.057$ & $14.780\pm0.125$ & $14.060\pm0.068$ & $12.977\pm0.031$ & Vega mag  \\
WISE $W1$ & $12.689\pm0.023$ & $13.817\pm0.025$ & $14.691\pm0.026$ & $13.870\pm0.024$ & $\cdots$ & Vega mag  \\
WISE $W2$ & $12.535\pm0.023$ & $13.915\pm0.028$ & $14.707\pm0.034$ & $14.019\pm0.028$ & $\cdots$ & Vega mag  \\
\textbf{Stellar Properties} & & & & &  &\\
Luminosity\tablenotemark{a} $L_{\ast}$ & $1681\pm10$ & $707^{+36}_{-24}$ & $378^{+14}_{-17}$ & $596\pm10$ & $1221^{+73}_{-39}$ & L$_{\odot}$  \\
Radius\tablenotemark{a} $R_{\ast}$ & $75\pm1$ & $43\pm1$ & $28\pm1$ &  $42\pm1$ & $69\pm1$ & R$_{\odot}$  \\
Distance\tablenotemark{a} $d_{\text{iso}}$ & $48.0\pm0.1$ & $49.1^{+1.3}_{-0.8}$ & $50.0^{+0.9}_{-1.2}$ & $51.2\pm0.1$ & $49.0^{+1.5}_{-0.8}$ & kpc  \\
Mass\tablenotemark{a} $M_{\odot}$ & $0.75\pm0.01$ & $0.81^{+0.02}_{-0.03}$ & $0.82^{+0.03}_{-0.02}$ & $0.77\pm0.01$ & $0.74\pm0.02$ & M$_{\odot}$  \\
Age\tablenotemark{a} $\tau$ & $13.3^{+0.1}_{-0.3}$ & $11.3^{+1.3}_{-0.9}$ & $11.9^{+1.1}_{-1.2}$ & $13.4^{+0.1}_{-0.2}$ & $12.3\pm0.9$ & Gyr  \\
Extinction\tablenotemark{a} $A_{V}$ & $0.171^{+0.004}_{-0.007}$ & $0.460^{+0.024}_{-0.011}$ & $0.155^{+0.007}_{-0.008}$ & $0.165^{+0.001}_{-0.002}$ & $0.446^{+0.012}_{-0.006}$ & mag  \\
Effective temperature\tablenotemark{a} $T_{\text{eff}}$ & $4247\pm1$ & $4520^{+26}_{-18}$ & $4761^{+18}_{-15}$ &    $4361^{+2}_{-1}$ & $4139^{+14}_{-13}$ & K  \\
Surface gravity\tablenotemark{a} $\log{g}$ & $0.55\pm0.01$ & $1.06\pm0.02$ & $1.43\pm0.02$ & $1.06\pm0.01$ & $0.64\pm0.02$ & cm s$^{-2}$  \\
Metallicity $[\text{Fe/H}]$ & $-2.32\pm0.36$ & $-2.15\pm0.38$ & $-2.42\pm0.23$ & $-1.65\pm0.35$    & $-1.83\pm0.46$ &  \\
Microturbulence $\xi$ & $2.91\pm0.50$ & $2.55\pm0.75$ & $2.45\pm0.62$ & $2.74\pm0.83$ & $3.31\pm0.91$ & km s$^{-1}$  \\
\hline
\hline
Property & J05160009-6207287 & J05224088-6951471 & J05242202-6945073 & J06411906-7016314 && Units \\
         &               LMC &               LMC &               LMC &               LMC &&   \\
\hline
\textbf{Photometric Properties} & & & & &&  \\
SkyMapper $g$ & $16.542\pm0.008$ & $17.440\pm0.017$ & $17.791\pm0.129$ & $16.776\pm0.023$ && AB mag   \\
SkyMapper $r$ & $15.770\pm0.008$ & $16.783\pm0.029$ & $17.237\pm0.009$ & $15.944\pm0.006$ && AB mag   \\
SkyMapper $i$ & $15.228\pm0.006$ & $16.418\pm0.016$ & $16.945\pm0.011$ & $15.320\pm0.007$ && AB mag   \\
SkyMapper $z$ & $14.987\pm0.007$ & $16.303\pm0.036$ & $17.020\pm0.091$ & $15.050\pm0.004$ && AB mag   \\
2MASS $J$ & $13.777\pm0.028$ & $15.026\pm0.052$ & $15.711\pm0.080$ & $13.789\pm0.029$ && Vega mag   \\
2MASS $H$ & $13.026\pm0.024$ & $14.340\pm0.061$ & $15.043\pm0.083$ & $13.015\pm0.026$ && Vega mag   \\
2MASS $K_{\text{s}}$ & $12.928\pm0.026$ & $14.368\pm0.093$ & $15.075\pm0.158$ & $12.834\pm0.035$ && Vega mag   \\
WISE $W1$ & $12.825\pm0.023$ & $14.109\pm0.058$ & $\cdots$ & $12.775\pm0.023$ && Vega mag   \\
WISE $W2$ & $12.835\pm0.022$ & $14.529\pm0.059$ & $\cdots$ & $12.817\pm0.023$ && Vega mag   \\
\textbf{Stellar Properties} & & & & &&  \\
Luminosity\tablenotemark{a} $L_{\ast}$ & $1455^{+55}_{-54}$ & $517\pm22$ & $279^{+14}_{-13}$ & $1555^{+18}_{-31}$ && L$_{\odot}$   \\
Radius\tablenotemark{a} $R_{\ast}$ & $73\pm1$ & $35\pm1$ & $23\pm1$ & $78\pm1$ && R$_{\odot}$   \\
Distance\tablenotemark{a} $d_{\text{iso}}$ & $49.6\pm1.0$ & $49.7^{+1.1}_{-1.2}$ & $49.4^{+1.2}_{-1.0}$ & $50.8^{+0.3}_{-0.5}$ && kpc   \\
Mass\tablenotemark{a} $M_{\odot}$ & $0.78^{+0.03}_{-0.02}$ & $0.80^{+0.03}_{-0.02}$ & $0.83\pm0.03$ & $0.76^{+0.02}_{-0.01}$ && M$_{\odot}$   \\
Age\tablenotemark{a} $\tau$ & $11.6^{+1.2}_{-1.1}$ & $11.9^{+1.1}_{-1.3}$ & $11.5^{+1.3}_{-1.1}$ & $12.8^{+0.5}_{-0.8}$ && Gyr   \\
Extinction\tablenotemark{a} $A_{V}$ & $0.063^{+0.023}_{-0.028}$ & $0.271^{+0.058}_{-0.039}$ & $0.038^{+0.048}_{-0.026}$ & $0.220^{+0.004}_{-0.006}$ && mag   \\
Effective temperature\tablenotemark{a} $T_{\text{eff}}$ & $4151\pm17$ & $4619^{+45}_{-23}$ & $4903^{+31}_{-49}$ & $4093^{+8}_{-7}$ && K   \\
Surface gravity\tablenotemark{a} $\log{g}$ & $0.60\pm0.02$ & $1.24\pm0.03$ & $1.61\pm0.03$ & $0.53\pm0.01$ && cm s$^{-2}$   \\
Metallicity $[\text{Fe/H}]$ & $-1.62\pm0.24$ & $-2.14\pm0.32$ & $-1.99\pm0.47$ & $-1.54\pm0.14$ &&    \\
Microturbulence $\xi$ & $3.25\pm0.54$ & $3.16\pm0.45$ & $2.38\pm0.78$ & $3.28\pm0.47$ && km s$^{-1}$
\enddata
\tablenotetext{a}{We report random uncertainties derived under the
unlikely assumption that the MIST isochrone grid perfectly reproduces
all stellar properties.  There are almost certainly larger systematic
uncertainties that we have not investigated, though the excellent
agreement between our analysis and the results from ASPCAP for the two
stars that are part of both our sample and the SDSS DR16 sample seems
to indicate that any systematic uncertainties in our analysis cannot be
too large.}
\end{deluxetable*} 
\end{longrotatetable} 

\begin{longrotatetable}
\begin{deluxetable*}{lcccccccccccccccccccc}
\tabletypesize{\tiny}
\tablewidth{10cm}
\tablecaption{Mean Elemental Abundances\label{chem_abundances}}
\tablecolumns{21}
\tablehead{
\colhead{Species} & 
\colhead{n} & \colhead{log($\epsilon_X$)} &
\colhead{[X/Fe]} & \colhead{$\sigma_{[\text{X/Fe}]}$} &
\colhead{n} & \colhead{log($\epsilon_X$)} &
\colhead{[X/Fe]} & \colhead{$\sigma_{[\text{X/Fe}]}$} &
\colhead{n} & \colhead{log($\epsilon_X$)} &
\colhead{[X/Fe]} & \colhead{$\sigma_{[\text{X/Fe}]}$} &
\colhead{n} & \colhead{log($\epsilon_X$)} &
\colhead{[X/Fe]} & \colhead{$\sigma_{[\text{X/Fe}]}$} &
\colhead{n} & \colhead{log($\epsilon_X$)} &
\colhead{[X/Fe]} & \colhead{$\sigma_{[\text{X/Fe}]}$} }
\startdata
\hline 
  & \multicolumn{4}{c}{J00251849-7140074} & \multicolumn{4}{c}{J00262394-7128549} & \multicolumn{4}{c}{J00263959-7122102} & \multicolumn{4}{c}{J00273753-7125114} & \multicolumn{4}{c}{J05121686-6517147} \\ 
 & \multicolumn{4}{c}{SMC} & \multicolumn{4}{c}{SMC} & \multicolumn{4}{c}{SMC} & \multicolumn{4}{c}{SMC} & \multicolumn{4}{c}{LMC} \\ \hline 
\ion{Na}{1} & $2$ & $4.187$ & $0.300$ & $0.264$ & $3$ & $3.974$ & $-0.222$ & $0.210$ & $2$ & $4.168$ & $0.550$ & $0.264$ & $2$ & $3.447$ & $-0.508$ & $0.254$ & $1$ & $4.166$ & $0.267$ & $0.100$  \\ 
\ion{Na}{1}$_{\rm{NLTE}}$ & $2$ & $4.072$ & $0.166$ & $\cdots$ & $3$ & $3.877$ & $-0.339$ & $\cdots$ & $2$ & $3.916$ & $0.278$ & $\cdots$ & $2$ & $3.329$ & $-0.646$ & $\cdots$ & $1$ & $4.123$ & $0.204$ & $\cdots$  \\ 
\ion{Mg}{1} & $2$ & $5.710$ & $0.493$ & $0.287$ & $2$ & $5.662$ & $0.136$ & $0.211$ & $1$ & $5.523$ & $0.575$ & $0.123$ & $2$ & $5.651$ & $0.366$ & $0.090$ & $2$ & $6.300$ & $1.071$ & $0.236$  \\ 
\ion{Si}{1} & $3$ & $5.864$ & $0.687$ & $0.174$ & $4$ & $6.027$ & $0.541$ & $0.023$ & $4$ & $6.204$ & $1.296$ & $0.090$ & $2$ & $5.757$ & $0.512$ & $0.029$ & $2$ & $5.835$ & $0.646$ & $0.216$  \\ 
\ion{K}{1} & $1$ & $3.569$ & $0.832$ & $0.192$ & $1$ & $3.237$ & $0.191$ & $0.151$ & $1$ & $3.679$ & $1.211$ & $0.174$ & $1$ & $3.224$ & $0.419$ & $0.099$ & $1$ & $4.793$ & $2.044$ & $0.356$  \\ 
\ion{K}{1}$_{\rm{NLTE}}$ & $1$ & $2.854$ & $0.157$ & $\cdots$ & $1$ & $2.932$ & $-0.074$ & $\cdots$ & $1$ & $2.938$ & $0.510$ & $\cdots$ & $1$ & $2.711$ & $-0.054$ & $\cdots$ & $1$ & $3.580$ & $0.871$ & $\cdots$  \\ 
\ion{Ca}{1} & $14$ & $4.581$ & $0.614$ & $0.132$ & $9$ & $4.383$ & $0.107$ & $0.150$ & $8$ & $4.547$ & $0.849$ & $0.096$ & $12$ & $4.429$ & $0.394$ & $0.083$ & $11$ & $4.958$ & $0.979$ & $0.298$  \\ 
\ion{Sc}{2} & $5$ & $0.736$ & $-0.071$ & $0.079$ & $4$ & $0.870$ & $-0.246$ & $0.138$ & $4$ & $0.899$ & $0.361$ & $0.069$ & $4$ & $0.989$ & $0.114$ & $0.072$ & $3$ & $1.257$ & $0.438$ & $0.225$  \\ 
\ion{Ti}{1} & $13$ & $3.265$ & $0.628$ & $0.164$ & $9$ & $3.199$ & $0.253$ & $0.165$ & $10$ & $3.363$ & $0.995$ & $0.176$ & $12$ & $3.184$ & $0.479$ & $0.115$ & $48$ & $4.364$ & $1.715$ & $0.371$  \\ 
\ion{Ti}{2} & $5$ & $3.307$ & $0.670$ & $0.125$ & $6$ & $3.313$ & $0.367$ & $0.203$ & $5$ & $3.178$ & $0.810$ & $0.169$ & $6$ & $2.878$ & $0.173$ & $0.106$ & $38$ & $3.322$ & $0.673$ & $0.176$  \\ 
\ion{Cr}{1} & $6$ & $3.632$ & $0.345$ & $0.289$ & $5$ & $3.262$ & $-0.334$ & $0.241$ & $2$ & $3.178$ & $0.160$ & $0.141$ & $5$ & $3.419$ & $0.064$ & $0.228$ & $4$ & $3.150$ & $-0.149$ & $0.363$  \\ 
\ion{Mn}{1} & $2$ & $3.032$ & $-0.055$ & $0.143$ & $4$ & $2.809$ & $-0.587$ & $0.212$ & $3$ & $2.865$ & $0.047$ & $0.144$ & $3$ & $2.778$ & $-0.377$ & $0.129$ & $2$ & $2.546$ & $-0.553$ & $0.710$  \\ 
\ion{Fe}{1} & $31$ & $5.294$ & $\cdots$ & $\cdots$  & $26$ & $5.222$ & $\cdots$ & $\cdots$  & $27$ & $5.190$ & $\cdots$ & $\cdots$  & $23$ & $4.998$ & $\cdots$ & $\cdots$  & $28$ & $5.478$ & $\cdots$ & $\cdots$   \\ 
\ion{Fe}{2} & $24$ & $5.163$ & $\cdots$ & $\cdots$ & $20$ & $5.464$ & $\cdots$ & $\cdots$ & $14$ & $4.898$ & $\cdots$ & $\cdots$ & $12$ & $5.229$ & $\cdots$ & $\cdots$ & $18$ & $5.175$ & $\cdots$ & $\cdots$  \\ 
\ion{Co}{1} & $3$ & $3.321$ & $0.714$ & $0.214$ & $2$ & $3.081$ & $0.165$ & $0.214$ & $2$ & $2.920$ & $0.582$ & $0.162$ & $1$ & $2.822$ & $0.147$ & $0.041$ & $5$ & $3.391$ & $0.772$ & $0.201$  \\ 
\ion{Ni}{1} & $14$ & $4.172$ & $0.305$ & $0.144$ & $7$ & $3.882$ & $-0.294$ & $0.117$ & $9$ & $4.043$ & $0.445$ & $0.119$ & $7$ & $3.911$ & $-0.024$ & $0.134$ & $7$ & $4.360$ & $0.481$ & $0.261$  \\ 
\ion{Cu}{1} & $1$ & $2.073$ & $0.226$ & $0.144$ & $\cdots$ & $\cdots$ & $\cdots$ & $\cdots$ & $\cdots$ & $\cdots$ & $\cdots$ & $\cdots$ & $1$ & $1.785$ & $-0.130$ & $0.068$ & $1$ & $3.255$ & $1.396$ & $0.413$  \\ 
\ion{Zn}{1} & $\cdots$ & $\cdots$ & $\cdots$ & $\cdots$ & $\cdots$ & $\cdots$ & $\cdots$ & $\cdots$ & $\cdots$ & $\cdots$ & $\cdots$ & $\cdots$ & $\cdots$ & $\cdots$ & $\cdots$ & $\cdots$ & $1$ & $2.772$ & $0.533$ & $0.121$  \\ 
\ion{Y}{2} & $2$ & $-0.215$ & $-0.092$ & $0.205$ & $3$ & $0.599$ & $0.413$ & $0.404$ & $3$ & $0.221$ & $0.613$ & $0.182$ & $3$ & $0.257$ & $0.312$ & $0.377$ & $2$ & $-0.105$ & $0.006$ & $0.123$  \\ 
\ion{Ba}{2} & $3$ & $-0.205$ & $-0.142$ & $0.264$ & $2$ & $0.508$ & $0.262$ & $0.292$ & $3$ & $-0.428$ & $-0.096$ & $0.201$ & $3$ & $-0.461$ & $-0.466$ & $0.161$ & $3$ & $0.798$ & $0.849$ & $0.292$  \\ 
\ion{La}{2} & $\cdots$ & $\cdots$ & $\cdots$ & $\cdots$ & $\cdots$ & $\cdots$ & $\cdots$ & $\cdots$ & $\cdots$ & $\cdots$ & $\cdots$ & $\cdots$ & $\cdots$ & $\cdots$ & $\cdots$ & $\cdots$ & $3$ & $0.153$ & $1.364$ & $0.278$  \\ 
\ion{Ce}{2} & $2$ & $-0.664$ & $0.089$ & $0.062$ & $3$ & $1.305$ & $1.749$ & $1.619$ & $2$ & $-0.355$ & $0.667$ & $0.280$ & $2$ & $-0.545$ & $0.140$ & $0.282$ & $3$ & $-0.118$ & $0.623$ & $0.588$  \\ 
\ion{Nd}{2} & $2$ & $-1.009$ & $-0.096$ & $0.188$ & $5$ & $-0.166$ & $0.438$ & $0.196$ & $3$ & $-0.602$ & $0.580$ & $0.198$ & $\cdots$ & $\cdots$ & $\cdots$ & $\cdots$ & $5$ & $0.370$ & $1.271$ & $0.434$  \\ 
\ion{Sm}{2} & $1$ & $-0.865$ & $0.518$ & $0.056$ & $2$ & $-0.409$ & $0.665$ & $0.275$ & $\cdots$ & $\cdots$ & $\cdots$ & $\cdots$ & $\cdots$ & $\cdots$ & $\cdots$ & $\cdots$ & $1$ & $-0.900$ & $0.471$ & $0.043$  \\ 
\ion{Gd}{2} & $1$ & $0.005$ & $1.258$ & $0.133$ & $2$ & $-0.132$ & $0.812$ & $0.266$ & $1$ & $-0.382$ & $1.140$ & $0.045$ & $\cdots$ & $\cdots$ & $\cdots$ & $\cdots$ & $2$ & $-0.579$ & $0.662$ & $0.122$  \\ 
\ion{Eu}{2} & $1$ & $-0.960$ & $0.853$ & $0.300$ & $1$ & $-0.500$ & $1.004$ & $0.300$ & $\cdots$ & $\cdots$ & $\cdots$ & $\cdots$ & $\cdots$ & $\cdots$ & $\cdots$ & $\cdots$ & $1$ & $-0.250$ & $1.551$ & $0.300$  \\ 
\hline 
 & \multicolumn{4}{c}{J05133509-7109322} & \multicolumn{4}{c}{J05141665-6454310} & \multicolumn{4}{c}{J05143154-6505189} & \multicolumn{4}{c}{J05150380-6647003} & \multicolumn{4}{c}{J05160009-6207287} \\ 
 & \multicolumn{4}{c}{LMC} & \multicolumn{4}{c}{LMC} & \multicolumn{4}{c}{LMC} & \multicolumn{4}{c}{LMC} & \multicolumn{4}{c}{LMC} \\ \hline 
\ion{Na}{1} & $3$ & $4.209$ & $0.144$ & $0.167$ & $2$ & $4.111$ & $0.311$ & $0.141$ & $2$ & $5.308$ & $0.736$ & $0.276$ & $2$ & $5.406$ & $1.018$ & $0.272$ & $2$ & $4.826$ & $0.223$ & $0.122$  \\ 
\ion{Na}{1}$_{\rm{NLTE}}$ & $3$ & $4.100$ & $0.015$ & $\cdots$ & $2$ & $4.053$ & $0.234$ & $\cdots$ & $2$ & $5.159$ & $0.567$ & $\cdots$ & $2$ & $5.237$ & $0.830$ & $\cdots$ & $2$ & $4.755$ & $0.133$ & $\cdots$  \\ 
\ion{Mg}{1} & $4$ & $5.868$ & $0.473$ & $0.230$ & $6$ & $5.660$ & $0.530$ & $0.136$ & $3$ & $6.116$ & $0.214$ & $0.419$ & $1$ & $5.836$ & $0.118$ & $0.229$ & $6$ & $5.862$ & $-0.071$ & $0.302$  \\ 
\ion{Si}{1} & $4$ & $6.187$ & $0.832$ & $0.081$ & $4$ & $5.993$ & $0.903$ & $0.092$ & $4$ & $6.834$ & $0.972$ & $0.100$ & $5$ & $6.361$ & $0.683$ & $0.079$ & $7$ & $6.575$ & $0.682$ & $0.094$  \\ 
\ion{K}{1} & $1$ & $3.338$ & $0.423$ & $0.200$ & $1$ & $5.854$ & $3.204$ & $0.194$ & $1$ & $4.358$ & $0.936$ & $0.325$ & $1$ & $3.886$ & $0.648$ & $0.300$ & $1$ & $3.482$ & $0.029$ & $0.205$  \\ 
\ion{K}{1}$_{\rm{NLTE}}$ & $1$ & $3.010$ & $0.135$ & $\cdots$ & $1$ & $3.850$ & $1.240$ & $\cdots$ & $1$ & $3.960$ & $0.578$ & $\cdots$ & $1$ & $2.979$ & $-0.219$ & $\cdots$ & $1$ & $2.578$ & $-0.835$ & $\cdots$  \\ 
\ion{Ca}{1} & $16$ & $4.615$ & $0.470$ & $0.204$ & $17$ & $4.253$ & $0.373$ & $0.136$ & $7$ & $5.342$ & $0.690$ & $0.321$ & $6$ & $5.054$ & $0.586$ & $0.258$ & $17$ & $4.620$ & $-0.063$ & $0.223$  \\ 
\ion{Sc}{2} & $5$ & $1.136$ & $0.151$ & $0.127$ & $3$ & $0.746$ & $0.026$ & $0.141$ & $5$ & $1.654$ & $0.162$ & $0.210$ & $5$ & $1.431$ & $0.123$ & $0.213$ & $5$ & $1.534$ & $0.011$ & $0.177$  \\ 
\ion{Ti}{1} & $43$ & $3.253$ & $0.438$ & $0.228$ & $35$ & $3.031$ & $0.481$ & $0.139$ & $24$ & $3.729$ & $0.407$ & $0.269$ & $19$ & $3.709$ & $0.571$ & $0.271$ & $35$ & $3.433$ & $0.080$ & $0.183$  \\ 
\ion{Ti}{2} & $35$ & $3.243$ & $0.428$ & $0.167$ & $44$ & $2.887$ & $0.337$ & $0.120$ & $23$ & $3.563$ & $0.241$ & $0.186$ & $26$ & $3.435$ & $0.297$ & $0.146$ & $35$ & $3.619$ & $0.266$ & $0.184$  \\ 
\ion{Cr}{1} & $13$ & $3.589$ & $0.124$ & $0.279$ & $10$ & $3.193$ & $-0.007$ & $0.151$ & $5$ & $4.129$ & $0.157$ & $0.399$ & $3$ & $3.677$ & $-0.111$ & $0.359$ & $15$ & $3.602$ & $-0.401$ & $0.309$  \\ 
\ion{Mn}{1} & $5$ & $2.965$ & $-0.300$ & $0.244$ & $5$ & $2.689$ & $-0.311$ & $0.157$ & $3$ & $2.912$ & $-0.860$ & $0.461$ & $3$ & $3.428$ & $-0.160$ & $0.239$ & $6$ & $3.570$ & $-0.233$ & $0.116$  \\ 
\ion{Fe}{1} & $79$ & $5.222$ & $\cdots$ & $\cdots$  & $88$ & $5.094$ & $\cdots$ & $\cdots$  & $36$ & $5.398$ & $\cdots$ & $\cdots$  & $29$ & $5.466$ & $\cdots$ & $\cdots$  & $37$ & $5.354$ & $\cdots$ & $\cdots$   \\ 
\ion{Fe}{2} & $28$ & $5.340$ & $\cdots$ & $\cdots$ & $26$ & $5.076$ & $\cdots$ & $\cdots$ & $24$ & $5.841$ & $\cdots$ & $\cdots$ & $23$ & $5.658$ & $\cdots$ & $\cdots$ & $26$ & $5.877$ & $\cdots$ & $\cdots$  \\ 
\ion{Co}{1} & $5$ & $3.216$ & $0.431$ & $0.223$ & $4$ & $3.157$ & $0.637$ & $0.342$ & $5$ & $3.985$ & $0.693$ & $0.128$ & $4$ & $3.715$ & $0.607$ & $0.190$ & $6$ & $3.574$ & $0.251$ & $0.049$  \\ 
\ion{Ni}{1} & $16$ & $4.105$ & $0.060$ & $0.177$ & $6$ & $3.851$ & $0.071$ & $0.173$ & $12$ & $4.678$ & $0.126$ & $0.240$ & $19$ & $4.533$ & $0.165$ & $0.164$ & $22$ & $4.367$ & $-0.216$ & $0.151$  \\ 
\ion{Cu}{1} & $1$ & $2.348$ & $0.323$ & $0.247$ & $\cdots$ & $\cdots$ & $\cdots$ & $\cdots$ & $1$ & $1.819$ & $-0.713$ & $0.154$ & $1$ & $4.115$ & $1.767$ & $0.421$ & $1$ & $1.906$ & $-0.657$ & $0.141$  \\ 
\ion{Zn}{1} & $2$ & $2.591$ & $0.186$ & $0.064$ & $2$ & $2.498$ & $0.358$ & $0.144$ & $2$ & $3.022$ & $0.110$ & $0.161$ & $2$ & $3.030$ & $0.302$ & $0.140$ & $2$ & $2.982$ & $0.039$ & $0.276$  \\ 
\ion{Y}{2} & $4$ & $-0.168$ & $-0.223$ & $0.135$ & $5$ & $-0.010$ & $0.200$ & $0.287$ & $3$ & $-0.026$ & $-0.588$ & $0.135$ & $4$ & $0.713$ & $0.335$ & $0.367$ & $3$ & $0.640$ & $0.047$ & $0.208$  \\ 
\ion{Ba}{2} & $3$ & $-0.055$ & $-0.170$ & $0.342$ & $3$ & $-0.477$ & $-0.327$ & $0.170$ & $3$ & $0.837$ & $0.215$ & $0.323$ & $3$ & $0.369$ & $-0.069$ & $0.297$ & $3$ & $0.012$ & $-0.641$ & $0.258$  \\ 
\ion{La}{2} & $2$ & $-1.020$ & $0.025$ & $0.252$ & $3$ & $-0.532$ & $0.778$ & $0.141$ & $3$ & $0.059$ & $0.597$ & $0.314$ & $3$ & $-0.121$ & $0.601$ & $0.248$ & $2$ & $-0.290$ & $0.217$ & $0.202$  \\ 
\ion{Ce}{2} & $5$ & $-0.602$ & $-0.027$ & $0.287$ & $2$ & $-0.706$ & $0.134$ & $0.308$ & $4$ & $0.148$ & $0.216$ & $0.258$ & $3$ & $-0.392$ & $-0.140$ & $0.352$ & $2$ & $0.053$ & $0.090$ & $0.133$  \\ 
\ion{Nd}{2} & $6$ & $-0.491$ & $0.244$ & $0.133$ & $5$ & $-0.828$ & $0.172$ & $0.138$ & $5$ & $0.077$ & $0.305$ & $0.198$ & $6$ & $0.079$ & $0.491$ & $0.322$ & $3$ & $0.014$ & $0.211$ & $0.138$  \\ 
\ion{Sm}{2} & $1$ & $-0.449$ & $0.756$ & $0.090$ & $1$ & $-1.124$ & $0.346$ & $0.062$ & $1$ & $0.316$ & $1.014$ & $0.204$ & $1$ & $-0.076$ & $0.806$ & $0.112$ & $2$ & $-0.135$ & $0.532$ & $0.277$  \\ 
\ion{Gd}{2} & $2$ & $-0.562$ & $0.513$ & $0.157$ & $3$ & $-0.363$ & $0.977$ & $0.249$ & $2$ & $-0.043$ & $0.525$ & $0.162$ & $1$ & $-0.040$ & $0.712$ & $0.087$ & $2$ & $0.451$ & $0.988$ & $0.277$  \\ 
\ion{Eu}{2} & $1$ & $-0.980$ & $0.655$ & $0.150$ & $2$ & $-1.500$ & $0.400$ & $0.300$ & $1$ & $-0.230$ & $0.898$ & $0.300$ & $1$ & $-0.260$ & $1.052$ & $0.150$ & $1$ & $-0.450$ & $0.647$ & $0.200$  \\ 
\hline 
 & \multicolumn{4}{c}{J05224088-6951471} & \multicolumn{4}{c}{J05242202-6945073} & \multicolumn{4}{c}{J06411906-7016314} \\ 
 & \multicolumn{4}{c}{LMC} & \multicolumn{4}{c}{LMC} & \multicolumn{4}{c}{LMC} \\ \hline 
\ion{Na}{1} & $2$ & $3.922$ & $-0.154$ & $0.317$ & $2$ & $4.172$ & $-0.051$ & $0.406$ & $2$ & $4.290$ & $-0.393$ & $0.151$ & &  &  & & &  &  & \\ 
\ion{Na}{1}$_{\rm{NLTE}}$ & $2$ & $3.383$ & $-0.670$ & $\cdots$ & $2$ & $3.664$ & $-0.560$ & $\cdots$ & $2$ & $4.275$ & $-0.428$ & $\cdots$ & &  &  & & &  &  & \\ 
\ion{Mg}{1} & $6$ & $5.625$ & $0.219$ & $0.164$ & $4$ & $5.993$ & $0.440$ & $0.185$ & $4$ & $5.850$ & $-0.163$ & $0.244$ & &  &  & & &  &  & \\ 
\ion{Si}{1} & $4$ & $5.748$ & $0.382$ & $0.080$ & $6$ & $5.902$ & $0.389$ & $0.088$ & $6$ & $6.153$ & $0.180$ & $0.114$ & &  &  & & &  &  & \\ 
\ion{K}{1} & $1$ & $2.638$ & $-0.430$ & $0.123$ & $1$ & $5.660$ & $1.257$ & $0.221$ & $1$ & $3.325$ & $-0.208$ & $0.165$ & &  &  & & &  &  & \\ 
\ion{K}{1}$_{\rm{NLTE}}$ & $1$ & $2.490$ & $-0.236$ & $\cdots$ & $1$ & $4.330$ & $0.776$ & $\cdots$ & $1$ & $3.111$ & $-0.382$ & $\cdots$ & &  &  & & &  &  & \\ 
\ion{Ca}{1} & $19$ & $4.337$ & $0.181$ & $0.129$ & $19$ & $4.576$ & $0.273$ & $0.135$ & $19$ & $4.535$ & $-0.228$ & $0.180$ & &  &  & & &  &  & \\ 
\ion{Sc}{2} & $5$ & $1.025$ & $0.029$ & $0.124$ & $5$ & $1.142$ & $-0.001$ & $0.121$ & $5$ & $1.531$ & $-0.072$ & $0.168$ & &  &  & & &  &  & \\ 
\ion{Ti}{1} & $38$ & $3.251$ & $0.425$ & $0.176$ & $27$ & $3.531$ & $0.558$ & $0.159$ & $36$ & $3.160$ & $-0.273$ & $0.140$ & &  &  & & &  &  & \\ 
\ion{Ti}{2} & $33$ & $3.035$ & $0.209$ & $0.105$ & $33$ & $3.171$ & $0.198$ & $0.149$ & $42$ & $3.552$ & $0.119$ & $0.181$ & &  &  & & &  &  & \\ 
\ion{Cr}{1} & $14$ & $3.331$ & $-0.145$ & $0.183$ & $11$ & $3.614$ & $-0.009$ & $0.187$ & $14$ & $3.477$ & $-0.606$ & $0.250$ & &  &  & & &  &  & \\ 
\ion{Mn}{1} & $5$ & $3.082$ & $-0.194$ & $0.184$ & $4$ & $2.657$ & $-0.766$ & $0.130$ & $5$ & $3.291$ & $-0.592$ & $0.051$ & &  &  & & &  &  & \\ 
\ion{Fe}{1} & $81$ & $5.119$ & $\cdots$ & $\cdots$  & $78$ & $5.430$ & $\cdots$ & $\cdots$  & $34$ & $5.451$ & $\cdots$ & $\cdots$  & &  &  & & &  &  & \\ 
\ion{Fe}{2} & $26$ & $5.306$ & $\cdots$ & $\cdots$ & $26$ & $5.453$ & $\cdots$ & $\cdots$ & $25$ & $5.954$ & $\cdots$ & $\cdots$ & &  &  & & &  &  & \\ 
\ion{Co}{1} & $3$ & $3.158$ & $0.362$ & $0.195$ & $3$ & $3.401$ & $0.458$ & $0.147$ & $5$ & $3.278$ & $-0.125$ & $0.070$ & &  &  & & &  &  & \\ 
\ion{Ni}{1} & $15$ & $4.073$ & $0.017$ & $0.160$ & $11$ & $4.452$ & $0.249$ & $0.203$ & $22$ & $4.329$ & $-0.334$ & $0.123$ & &  &  & & &  &  & \\ 
\ion{Cu}{1} & $\cdots$ & $\cdots$ & $\cdots$ & $\cdots$ & $\cdots$ & $\cdots$ & $\cdots$ & $\cdots$ & $1$ & $1.975$ & $-0.668$ & $0.138$ & &  &  & & &  &  & \\ 
\ion{Zn}{1} & $2$ & $2.569$ & $0.153$ & $0.143$ & $2$ & $2.730$ & $0.167$ & $0.131$ & $2$ & $2.671$ & $-0.352$ & $0.094$ & &  &  & & &  &  & \\ 
\ion{Y}{2} & $5$ & $0.229$ & $0.163$ & $0.117$ & $2$ & $0.207$ & $-0.006$ & $0.177$ & $3$ & $0.285$ & $-0.388$ & $0.122$ & &  &  & & &  &  & \\ 
\ion{Ba}{2} & $3$ & $0.359$ & $0.233$ & $0.251$ & $3$ & $-0.031$ & $-0.304$ & $0.220$ & $3$ & $0.049$ & $-0.684$ & $0.236$ & &  &  & & &  &  & \\ 
\ion{La}{2} & $3$ & $0.113$ & $1.147$ & $0.100$ & $3$ & $-0.412$ & $0.475$ & $0.216$ & $2$ & $-0.478$ & $-0.051$ & $0.216$ & &  &  & & &  &  & \\ 
\ion{Ce}{2} & $5$ & $-0.106$ & $0.458$ & $0.227$ & $2$ & $0.433$ & $0.850$ & $0.256$ & $2$ & $0.084$ & $0.041$ & $0.143$ & &  &  & & &  &  & \\ 
\ion{Nd}{2} & $6$ & $0.419$ & $1.143$ & $0.246$ & $5$ & $-0.052$ & $0.525$ & $0.147$ & $3$ & $0.198$ & $0.315$ & $0.135$ & &  &  & & &  &  & \\ 
\ion{Sm}{2} & $2$ & $0.055$ & $1.249$ & $0.162$ & $2$ & $-0.421$ & $0.626$ & $0.125$ & $2$ & $-0.394$ & $0.193$ & $0.099$ & &  &  & & &  &  & \\ 
\ion{Gd}{2} & $3$ & $0.394$ & $1.458$ & $0.202$ & $1$ & $-0.106$ & $0.811$ & $0.093$ & $1$ & $0.044$ & $0.501$ & $0.094$ & &  &  & & &  &  & \\ 
\ion{Eu}{2} & $1$ & $-0.410$ & $1.214$ & $0.150$ & $1$ & $-0.990$ & $0.487$ & $0.300$ & $1$ & $-0.520$ & $0.497$ & $0.200$ & &  &  & & &  &  & \\ 
\enddata
\end{deluxetable*}
\end{longrotatetable}

\end{document}